\documentclass[aps,amssymb,nofootinbib]{revtex4}
\usepackage[T1]{fontenc}
\usepackage[latin9]{inputenc}
\usepackage{color}
\usepackage{amsmath}
\usepackage{amssymb}
\usepackage{wasysym}
\usepackage{graphicx}
\usepackage{esint}
\usepackage[unicode=true,
 bookmarks=false,
 breaklinks=false,pdfborder={0 0 1},backref=section,colorlinks=false]
 {hyperref}

\makeatletter

\providecommand{\tabularnewline}{\\}

\@ifundefined{textcolor}{}
{%
 \definecolor{BLACK}{gray}{0}
 \definecolor{WHITE}{gray}{1}
 \definecolor{RED}{rgb}{1,0,0}
 \definecolor{GREEN}{rgb}{0,1,0}
 \definecolor{BLUE}{rgb}{0,0,1}
 \definecolor{CYAN}{cmyk}{1,0,0,0}
 \definecolor{MAGENTA}{cmyk}{0,1,0,0}
 \definecolor{YELLOW}{cmyk}{0,0,1,0}
}

\@ifundefined{definecolor}{\@ifundefined{definecolor}
 {\usepackage{color}}{}
}{}
\@ifundefined{definecolor}{\@ifundefined{definecolor}
 {\@ifundefined{definecolor}
 {\usepackage{color}}{}
}{}
}{}

\makeatother

\begin{document}

\title{Surfing gravitational waves: can bigravity survive growing tensor
modes?}

\author{Luca Amendola$^{1}$, Frank Koennig$^{1}$, Matteo Martinelli$^{1}$,
Valeria Pettorino$^{1}$, Miguel Zumalacarregui$^{1}$}

\affiliation{$^{1}$Institut für Theoretische Physik, Ruprecht-Karls-Universität
Heidelberg, Philosophenweg 16, 69120 Heidelberg, Germany}
\begin{abstract}
The theory of bigravity offers one of the simplest possibilities to
describe a massive graviton while having self-accelerating cosmological
solutions without a cosmological constant. However, it has been shown
recently that bigravity is affected by early-time fast growing modes
on the tensor sector. Here we argue that we can only trust the linear
analysis up to when perturbations are in the linear regime and use
a cut-off to stop the growing of the metric perturbations. This analysis,
although more consistent, still leads to growing tensor modes that
are unacceptably large for the theory to be compatible with measurements
of the cosmic microwave background (CMB), both in temperature and
polarization spectra. In order to suppress the growing modes and make
the model compatible with CMB spectra, we find it necessary to either
fine-tune the initial conditions, modify the theory or set the cut-off
for the tensor perturbations of the second metric much lower than
unity. Initial conditions such that the growing mode is sufficiently
suppresed can be achieved in scenarios in which inflation ends at
the GeV scale. 
\end{abstract}
\maketitle

\section{Introduction}
Evidence from an increasing number of cosmological observables favours an accelerating
universe at late times \cite{Perlmutter1998,Riess1998,Anderson:14,Betoule:2014,Beutler:2013yhm,Samushia:2013yga,2012MNRAS.427..146H,2015arXiv150201590P, 2015arXiv150201589P}.
This era of accelerated expansion may be due to novel gravitational
physics, which will be tested by ongoing and future experiments \cite{Amendola:2012ys}.
This possibility has triggered vigorous interest in alternative theories
of gravity \cite{Amendola2010,Clifton:2011jh, 2015arXiv150201590P}. Any modification of
gravity requires new degrees of freedom (dof). Since the theory of
a massless graviton is unique, new dofs are often gained by adding
new fields. The simplest possibility is the addition of a scalar field,
typically resulting in theories belonging to the Horndeski class \cite{Horndeski:1974,Deffayet:2011gz}
or beyond \cite{Zumalacarregui:2013pma,Gleyzes:2014dya}.

Formulating a theory of massive gravity has been
a long standing problem in theoretical physics due to the difficulties
to incorporate the right degrees of freedom. The linear Fierz-Pauli
theory had been developed long time ago \cite{Fierz:1939ix}, but
until recently all non-linear completions introduced the so called
Bouleware-Deser (BD) ghost \cite{BoulwareDeser1972}, an extra dof
that makes the theory not viable. Despite the difficulties, a class
of healthy theories has been recently identified \cite{2011PhRvL.106w1101D}
in which a specific choice of the potential terms makes the theory
ghost-free \cite{2012PhRvL.108d1101H}. All these theories of massive
gravity describe an interaction of two tensor fields in which the
second one, the so called \emph{reference metric}, is fixed. While
massive gravity only allows static solutions on homogeneous backgrounds
\cite{2011PhRvD..84l4046D}, a bimetric theory with a dynamical reference
metric does not introduce the BD ghost and describes dynamical cosmologies
\cite{2012JHEP...02..126H,2012JHEP...04..123H,2012JHEP...02..026H}
(see also the reviews \cite{2012RvMP...84..671H,2014arXiv1401.4173D}).
Cosmological solutions in these bimetric theories often allow for
self-acceleration without the introduction of a cosmological constant
\cite{1475-7516-2012-03-042} and were successfully compared to observations
at background level \cite{2013JHEP...03..099A,1475-7516-2014-03-029,1475-7516-2012-03-042}.

Many bigravity theories are however
affected by gradient instabilities in their scalar sector, as has
been shown by studies of the linear perturbations \cite{Konnig:2014dna,Comelli:2012db,DeFelice:2014nja}
(see Refs. \cite{Comelli:2012db,Berg:2012kn} for derivations of the
equations Refs. \cite{2014arXiv1404.4061S,Konnig:2014xva,Lagos:2014lca,Enander:2015vja,Enander:2013kza}
for discussion of their dynamics). Stable evolution can be achieved
only in a two parameters class of models known as Infinite-Branch
Bigravity (IBB) \cite{Konnig:2014xva}. In IBB, the reference metric
(in keeping with common usage, we keep referring to the second metric
as reference metric even if in reality is dynamical; we also use the
notation $f$-metric) is contracting during the radiation and most
of the matter era, until it undergoes a bounce at low redshift and
begins to expand, coinciding with the onset of accelerated expansion
in the physical metric without the need for a cosmological constant.
The early time contraction of the reference metric makes tensor perturbations
grow with time in IBB theories, as it was first shown in Ref. \cite{Cusin:2014psa,Lagos:2014lca}
(see also \cite{Dubovsky:2009xk,Comelli:2012db} for modified tensor
perturbation equations). This growing mode couples to the physical
metric and severely modifies its dynamics, leading to observable consequences.

In this paper we will
investigate the effects of these large tensor perturbations on the
Cosmic Microwave Background (CMB) and possible mechanisms to make
the theory compatible with current observations. The perturbations
in the reference metric grow very fast and rapidly become non-linear.
At this point we will assume that tensor perturbations stabilize,
modeling this effect by introducing a cut-off in the perturbations
of the reference metric. Despite this treatment, the tensor growing
mode significantly affects the evolution of the physical metric, and
the consequences can be seen as an enhancement of both temperature
and polarization spectra on low multipoles. These effects cannot
be sufficiently reduced by varying the bigravity or other cosmological
parameters: making the theory viable requires either fine tuning of
the initial conditions, lowering the cut-off or modifying the theory.
As it will be shown below, sufficient suppression of the growing mode
can be achieved by an inflationary mechanism that produces Hubble-scale
tensor perturbations at an energy scale of order few GeV.

\section{Bigravity}

\label{sec:bigravity}

We start with the action of the form \cite{2012JHEP...02..126H} 
\begin{eqnarray}
S & = & -\dfrac{M_{g}^{2}}{2}\int d^{4}x\sqrt{-g}R(g)-\dfrac{M_{f}^{2}}{2}\int d^{4}x\sqrt{-f}R(f)\\
 & + & m^{2}M_{g}^{2}\int d^{4}x\sqrt{-g}\sum_{n=0}^{4}\beta_{n}e_{n}(X)+\int d^{4}x\sqrt{-g}\mathcal{L}_{m}\nonumber 
\end{eqnarray}
where $e_{n}(X)$ are the elementary symmetric polynomials of the
eigenvalues of the matrices $X_{\gamma}^{\alpha}\equiv\sqrt{g^{\alpha\beta}f_{\beta\gamma}}$,
$M_{g}$ and $M_{f}$ are the Planck masses for $g_{\mu\nu}$ and
$f_{\mu\nu}$, respectively, $m$ is the mass scale of the graviton,
$\beta_{n}$ are arbitrary constants and $\mathcal{L}_{m}=\mathcal{L}_{m}(g,\psi)$
is the matter Lagrangian. Throughout the paper we will use a mostly
plus metric signature convention and natural units in which the speed
of light $c$ is set to one.

Here $g_{\mu\nu}$ is the standard metric coupled to matter fields
in the $\mathcal{L}_{m}$ Lagrangian, while $f_{\mu\nu}$ is an additional
dynamical tensor field. In the following we express masses in units
of the Planck mass $M_{g}$ and the mass parameter $m^{2}$ will be
absorbed into the parameters $\beta_{n}$. Varying the action with
respect to $g{}_{\mu\nu}$, one obtains the following equations of
motion: 
\begin{equation}
G_{\mu\nu}+\dfrac{1}{2}\sum_{n=0}^{3}(-1)^{n}m^{2}\beta_{n}\left[g_{\mu\lambda}Y_{(n)\nu}^{\lambda}(X)+g_{\nu\lambda}Y_{(n)\mu}^{\lambda}(X)\right]=T_{\mu\nu}\label{eq:eeg}
\end{equation}
where $G_{\mu\nu}$ is Einstein's tensor, and the expressions $Y_{(n)\nu}^{\lambda}(X)$
are defined as 
\begin{align}
Y_{(0)} & =I,\\
Y_{(1)} & =X-I[X],\\
Y_{(2)} & =X^{2}-X[X]+\dfrac{1}{2}I\left([X]^{2}-[X^{2}]\right)\\
Y_{(3)} & =X^{3}-X^{2}[X]+\dfrac{1}{2}X\left([X]^{2}-[X^{2}]\right))\nonumber \\
 & -\dfrac{1}{6}I\left([X]^{3}-3[X][X^{2}]+2[X^{3}]\right)
\end{align}
where $I$ is the identity matrix and $[...]$ is the trace operator.

Varying the action with respect to $f{}_{\mu\nu}$ we get 
\begin{equation}
\bar{G}_{\mu\nu}+\sum_{n=0}^{3}\frac{(-1)^{n}m^{2}\beta_{4-n}}{2M_{f}^{2}}\left[f_{\mu\lambda}Y_{(n)\nu}^{\lambda}(X^{-1})+f_{\nu\lambda}Y_{(n)\mu}^{\lambda}(X^{-1})\right]=0\label{eq:eef}
\end{equation}
where the overbar indicates $f{}_{\mu\nu}$ curvatures. Notice that
$\beta_{0}$ acts as a pure cosmological constant, which is however
not required to satisfy current observations. Finally, the rescaling
$f\rightarrow M_{f}^{-2}f$, $\beta_{n}\rightarrow M_{f}^{n}\beta_{n}$
allows us to assume $M_{f}=1$ in the following (see \cite{2012JCAP...12..021B}).
Additionally, from now on we absorb the graviton mass $m$ into the
constants $\beta_{i}$.

We assume now a cosmological spatially flat FRW metric: 
\begin{equation}
ds^{2}=a^{2}(\tau)\left(-d\tau^{2}+dx_{i}dx^{i}\right)
\end{equation}
where $\tau$ represents the conformal time and a dot will represent
the derivative with respect to it. The second metric is chosen also
in a spatially FRW form 
\begin{equation}
ds_{f}^{2}=-\left[\dot{b}(\tau)^{2}/\mathcal{H}^{2}(t)\right]d\tau^{2}+b(\tau)^{2}dx_{i}dx^{i}
\end{equation}
where $\mathcal{H}\equiv\dot{a}/a$ is the conformal Hubble function
and $b(\tau)$ is the `scale' factor associated with the second metric
$f$. This form of the metric $f_{\mu\nu}$ ensures that the equations
satisfy the Bianchi constraints (see e.g. \cite{2012JHEP...02..026H}).

The background equations for the two metrics have been obtained and
discussed at length in several papers \cite{2012JHEP...03..067C,2013JHEP...03..099A,Konnig:2014xva,1475-7516-2012-03-042}.
Here we summarize the main properties in the notation of \cite{Konnig:2014xva}.
Defining $r(\tau)\equiv b(\tau)/a(\tau)$ as the ratio of the two
scale factors, the background equations can be conveniently written
as a first order system of two equations for $r(t)$ and $\mathcal{H}$:
\begin{align}
2\mathcal{H}'\mathcal{H}+\mathcal{H}^{2} & =a^{2}(B_{0}+B_{2}r'-w_{tot}\rho_{tot}),\\
r' & =\frac{3rB_{1}\Omega_{tot}(1+w_{tot})}{\beta_{1}-3\beta_{3}r^{2}-2\beta_{4}r^{3}+3B_{2}r^{2}},\label{eq:rprime}
\end{align}
where the prime denotes derivative with respect to $N\equiv\log a$
\cite{1475-7516-2014-03-029,2013JHEP...03..099A}, $w_{tot}$ denotes
the equation of state (EOS) corresponding to the total density parameter
$\Omega_{tot}$ and the functions $B_{0}(\tau),B_{1}(\tau),B_{2}(\tau)$
are related to the $\beta_{i}$ and $r(\tau)$ as follows: 
\begin{align}
B_{0}(\tau) & =\beta_{0}+3\beta_{1}r+3\beta_{2}r^{2}+\beta_{3}r^{3},\\
B_{1}(\tau) & =\beta_{1}+3\beta_{2}r+3\beta_{3}r^{2}+\beta_{4}r^{3},\\
B_{2}(\tau) & =\beta_{1}+2\beta_{2}r+\beta_{3}r^{2}.
\end{align}
For simplicity the time dependence of $B_{0,1,2}$ will be understood
from now on. The Friedmann equation (i.e. the $(0,0)$ component of
Eq. (\ref{eq:eeg})) gives 
\begin{equation}
3\mathcal{H}^{2}=a^{2}(\rho_{tot}+B_{0})\,,
\end{equation}
and by combining with the $(0,0)$ component of Eq. (\ref{eq:eef})
we obtain a useful relation between $\mathcal{H}$ and the ratio $r(\tau)$:
\begin{equation}
\mathcal{H}^{2}=\frac{a^{2}B_{1}}{3r}.\label{eq:hubble}
\end{equation}
Finally, the combination of the last two expressions for $\mathcal{H}$
provide $\Omega_{tot}(\tau)=1-\frac{B_{0}}{B_{1}}r(\tau)$ which can
be inserted in Eq. (\ref{eq:rprime}) to produce a closed differential
equation for $r(\tau)$ alone.

The behavior of the background solutions depends on the choice of
the $\beta_{i}$ constants and on the initial value of $r$. We denote
solutions with the same $\beta_{i}$ but different initial conditions
as \emph{branches} of the same theory. In Ref. \cite{1475-7516-2014-03-029}
it was shown that for each choice of $\beta_{i}$ only two branches
exist that agree with a standard cosmological early time evolution
(like a matter dominated era at early times) and allow for physical
solutions (e.g. $\rho$, $\mathcal{H}>0$). In the first branch, $r$
evolves from $r=0$ to a de Sitter point at a finite value $r_{c}>0$.
These branches, however, suffer from scalar instabilities \cite{Konnig:2014dna}.
Only choosing $\beta_{2}=\beta_{3}=0$ and the second type of branches
in which $r$ evolves from $r\rightarrow\infty$ in the asymptotic
past towards a de Sitter point at a constant $r_{c}>0$, allows for
stable scalar perturbations and is compatible with background data
\cite{Konnig:2014xva}: we dubbed this case infinite-branch bigravity
(IBB). Note that even though an additional non-vanishing effective
cosmological constant $\beta_{0}$ is viable, we assume $\beta_{0}=0$
since it would not affect the early-time evolution and is not required
in order to fit observational data (see \cite{2013JHEP...03..099A,1475-7516-2014-03-029}).
From now on we restrict ourselves to IBB, in which only $\beta_{1}$
and $\beta_{4}$ are non-zero. This choice avoids introducing an explicit
cosmological constant, which would make the entire bigravity model
somewhat less appealing.

As shown in \cite{Konnig:2014xva}, IBB models have to satisfy $0<\beta_{4}<2\beta_{1}$
in order to get an initial value of $r$ on the infinite branch. In
particular, it was found that the best fit model occurs for $\beta_{1}=0.48$
and $\beta_{4}=0.94$: from now on we refer to this choice as the
reference IBB model. We then have 
\begin{equation}
\mathcal{H}^{2}=\frac{a^{2}(\beta_{1}+\beta_{4}r^{3})}{3r}.
\end{equation}
Here we derive the early time behaviour of the background evolution
for later use. 
(corresponding to early time in IBB), Eq. (\ref{eq:rprime}) for IBB
reduces to 
\begin{equation}
r'\sim-\frac{3}{2}(1+w_{tot})r\label{eq:early}
\end{equation}
so that for $w_{tot}={\rm {const}}$ (i.e., in radiation or matter
dominated epochs) one has 
\begin{equation}
r\sim a^{-3(1+w_{tot})/2}\,,
\end{equation}
and the $f_{\mu\nu}$ scale factor $b(\tau)=r(\tau)a(\tau)$ goes
as: 
\begin{equation}
b\sim a^{-(1+3w_{tot})/2}\,.
\end{equation}
The scale factor $b(\tau)$ therefore contracts instead of expanding
as long as $w_{tot}>-1/3$. Moreover, in the same approximation, 
\begin{equation}
\mathcal{H}^{2}\approx\frac{a^{2}\beta_{4}r^{2}}{3}.
\end{equation}

It is useful to derive an approximated estimate for the $b(\tau)$
bounce epoch. The bounce occurs $b'=0\;\Leftrightarrow\; r'\Big|_{r_{b}}=-r_{b}$.
If we assume a bounce after the radiation epoch, then the ratio at
the bounce has to satisfy 
\begin{equation}
4\beta_{1}-6\beta_{1}r_{b}^{2}+\beta_{4}r_{b}^{3}=0.\label{eq:condition_rbounce}
\end{equation}
From comparisons with observational data, we know that the best fit
is close to $\beta_{4}\approx2\beta_{1}$ which leads to 
\begin{equation}
r_{b}\simeq1+\sqrt{3}\,.
\end{equation}
It is also useful, for future purposes, to take note of the approximate
observational relation between $\beta_{1}$ and $\beta_{4}$: 
\begin{equation}
\beta_{4}=\beta_{1}^{2}\frac{3\left(1-\Omega_{tot0}\right)-\beta_{1}^{2}}{\left(1-\Omega_{tot0}\right)^{3}}=\frac{3\beta_{1}^{2}}{\left(\Omega_{tot0}-1\right)^{2}}+\mathcal{O}(\beta_{1}^{4})\,\,;\label{eq:deg_curve}
\end{equation}
This relation approximately corresponds to the degenerate line between
$\beta_{1}$ and $\beta_{4}$ for a flat universe, when fitting data
sets such as supernovae, CMB and baryonic acoustic oscillation data
\cite{Konnig:2014xva}. Moreover, we require $\beta_{1}\lesssim0.5$,
to ensure that the solution of Eq. (\ref{eq:hubble}) at present time
lies on the infinite branch.

\section{Tensor perturbations}

\label{sec:tensor_perturb}

\begin{figure}
\begin{centering}
\includegraphics[width=0.5\textwidth]{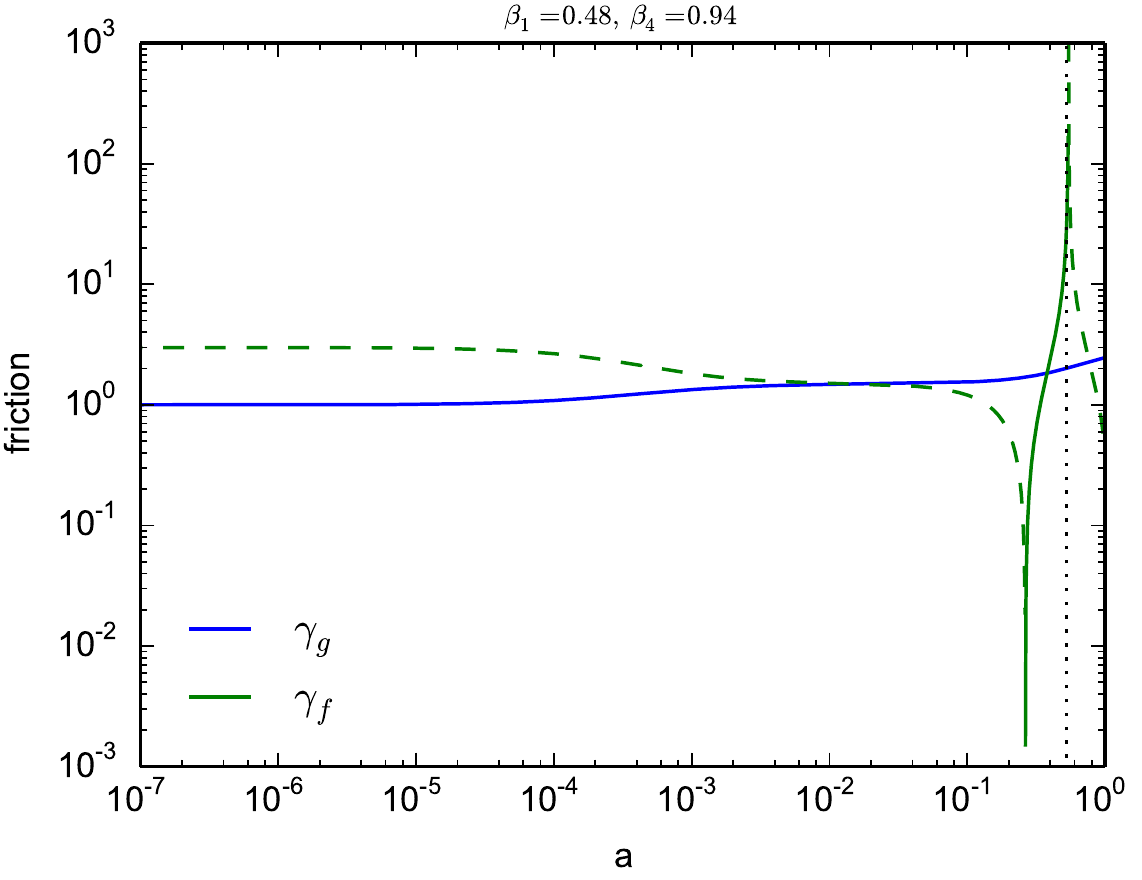}\includegraphics[width=0.5\textwidth]{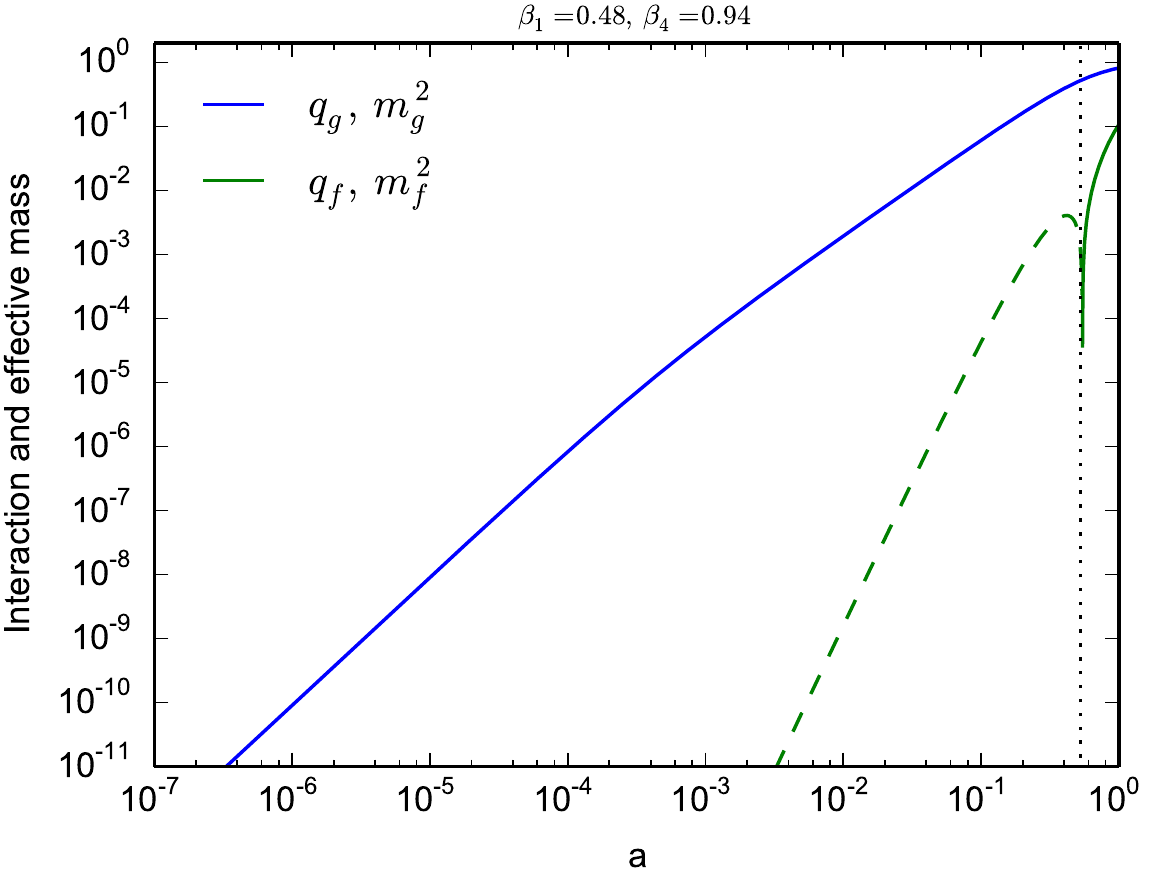}
\includegraphics[width=0.5\textwidth]{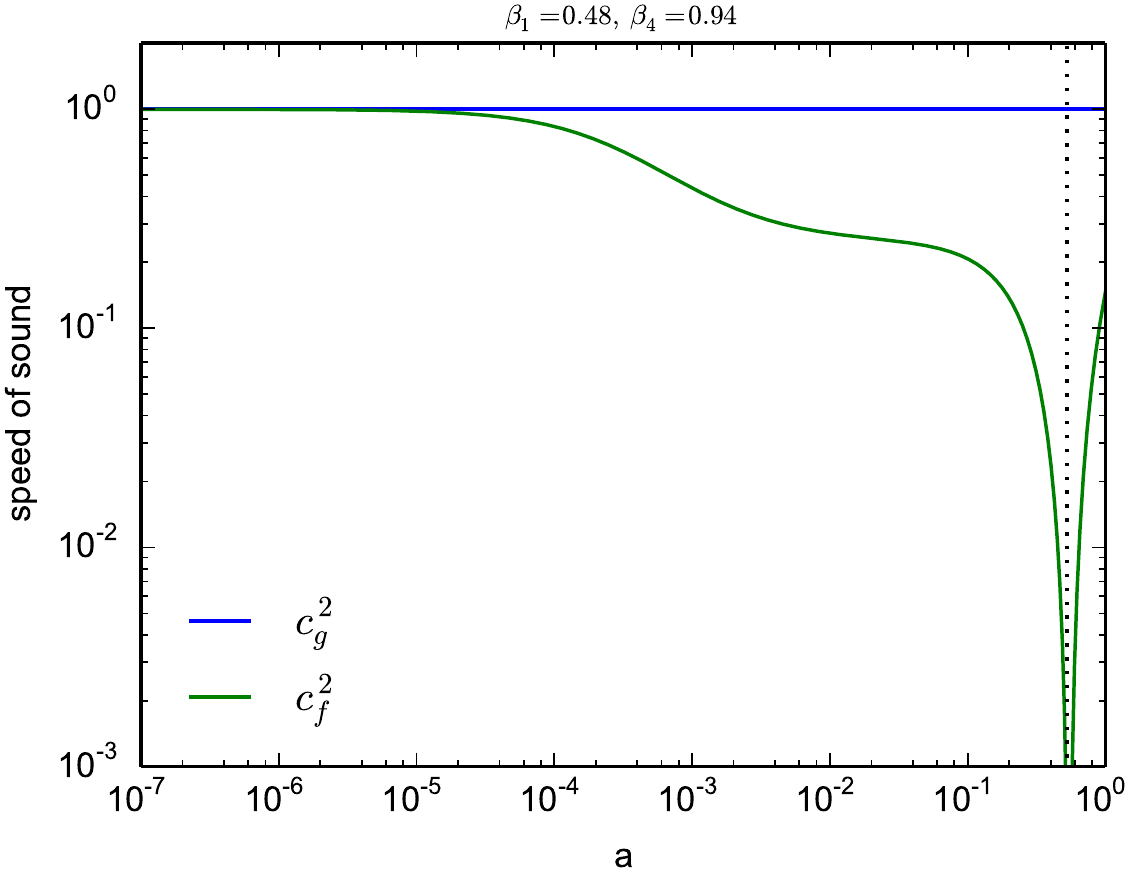}\includegraphics[width=0.5\textwidth]{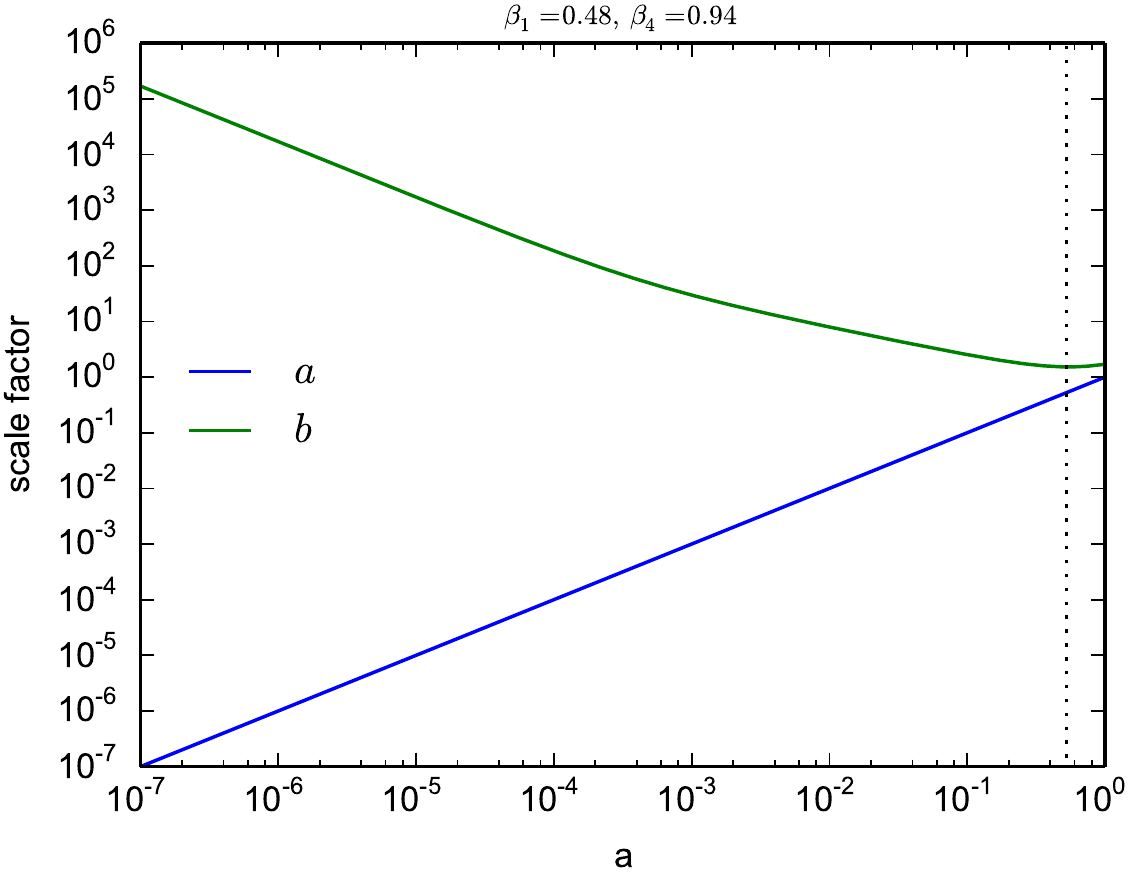} 
\par\end{centering}

\protect\protect\protect\protect\caption{Coefficients of the tensor equations for IBB (\ref{coeff1}-\ref{coeff4}).
Solid/dashed lines indicate positive/negative values and the vertical
dotted line marks the bounce of the reference metric. Note that the
friction term in the $f$-metric is negative at early times, when
the $f$-metric is contracting. Note also that the coupling and effective
mass terms are equal, and the ones corresponding to the $f$ metric
are very suppressed at early times. The bottom-right panel shows the
evolution of the two scale factors. \label{fig:tensor_coefficients} }
\end{figure}

As we are interested on the effect of bigravity on gravitational waves,
we now proceed with writing the tensor perturbation equations \cite{Cusin:2014psa,Comelli:2012db,Lagos:2014lca}.
For the perturbed part of the metrics we adopt the transverse-traceless
(TT) gauge, i.e. we select a transverse wave propagating along the
$z$ direction. Then, the tensor metric perturbations are given by:
\begin{equation}
h_{g(ij)}=\left(\begin{array}{ccc}
h_{g(+)} & h_{g(\times)} & 0\\
h_{g(\times)} & -h_{g(+)} & 0\\
0 & 0 & 0
\end{array}\right)
\end{equation}
\textit{\emph{and similarly for the tensor modes of the $f$ metric.}}\textit{
}\textit{\emph{We then obtain the following equations for both components
(suppressing the subscripts $+,\times$) 
\begin{equation}
h_{n}^{\prime\prime}+\gamma_{n}h_{n}^{\prime}+\left(m_{n}^{2}+c_{n}^{2}\mathcal{H}^{-2}k^{2}\right)h_{n}=q_{n}h_{m}\,,\label{eq:gweq}
\end{equation}
where the indices $n\not=m$ refer to $g$-metric and $f$-metric,
respectively; we have then 
\begin{align}
\gamma_{g} & =2+\frac{\mathcal{H}'}{\mathcal{H}},\quad\,\,\,\,\,\,\,\,\,\,\,\,\,\,\gamma_{f}=\frac{2r^{2}+3r'^{2}+r\left(4r'-r''\right)}{r\left(r'+r\right)}+\frac{\mathcal{H}'}{\mathcal{H}}\,\,;\label{coeff1}\\
m_{g}^{2} & =\mathcal{H}^{-2}a^{2}Br,\,\,\,\,\,\,\quad m_{f}^{2}=\frac{\left(r'+r\right)}{\mathcal{H}^{2}r^{2}}a^{2}B\,\,;\\
c_{g}^{2} & =1,\quad\,\,\,\,\,\,\,\,\,\,\,\,\,\,\,\,\,\,\,\,\,\,\,\,\,\,\,\,\,\, c_{f}^{2}=\frac{\left(r'+r\right)^{2}}{r^{2}}\,\,;\\
q_{g} & =\mathcal{H}^{-2}a^{2}Br,\quad\,\,\,\,\,\,\,\, q_{f}=\frac{\left(r'+r\right)}{\mathcal{H}^{2}r^{2}}a^{2}B\,\,,\label{coeff4}
\end{align}
and where: 
\begin{equation}
B\equiv\beta_{1}+\beta_{3}r^{2}+r\left(2\beta_{2}+\beta_{3}r'\right)+\beta_{2}r'.
\end{equation}
These equations are equivalent to the ones in Refs. \cite{Comelli:2012db,Lagos:2014lca}.
In IBB (i.e. for $\beta_{0}=\beta_{2}=\beta_{3}=0$), $B$ is simply
given by $\beta_{1}$.}} The coefficients (\ref{coeff1})-(\ref{coeff4})
for the two tensor equations are plotted in Fig.\ref{fig:tensor_coefficients}
as a function of redshift, for the choice $\beta_{1}=0.48$ and $\beta_{2}$
= 0.94. For this reference model, when considering both matter and
radiation, the bounce happens at a redshift $z_{b}\simeq0.9$, with
a corresponding $r_{b}\simeq2.8$.

Let us anticipate here an important feature of these equations. As
it will be shown below (see also Ref. \cite{Cusin:2014psa}), the
equation for $h_{f}$ is unstable at early times since its friction
term is negative as long as the scale factor $b(t)$ is collapsing
instead of expanding. The fast growth of $h_{f}$ will then drive
a fast growth of $h_{g}$ as well, through the coupling term. However,
in the limit $r\rightarrow\infty$, the coupling coefficient $q_{g}$
becomes 
\begin{equation}
q_{g}=\frac{a^{2}r\beta_{1}}{\mathcal{H}^{2}}\xrightarrow{r\rightarrow\infty}3\frac{\beta_{1}}{\beta_{4}r} \,\, , \label{eq:coupling_early}
\end{equation}
and is therefore relatively small for large $r$. For the reference
IBB model we have $\beta_{1}/\beta_{4}\approx0.5$; more in general,
according to Eq. (\ref{eq:deg_curve}), $\beta_{1}/\beta_{4}\approx1/3\beta_{1}$
to within factors of order unity, and therefore since $\beta_{1}<0.5,$
we have the lower bound 
\begin{equation}
q_{g}\approx\frac{1}{\beta_{1}r}\ge\frac{1}{r} \, \, . \label{eq:couplqg}
\end{equation}
At recombination, for instance, we have $r\approx10^{4}$ in IBB so
that one needs a $h_{f}$ roughly $10^{4}$ times bigger than $h_{g}$
before the coupling term $q_{g}h_{f}$ becomes comparable to the $h_{g}$
terms in Eq. (\ref{eq:gweq}) and it starts driving the evolution
of $h_{g}$. This means that, in principle, a growing mode in $h_{f}$
will take some time before affecting $h_{g}$. Whether this is enough
to spoil the physical metric, is what we are going to test below.

In the following subsections we discuss more in detail the time behavior
of $h_{g},h_{f}$ during the inflationary, radiation and matter eras.

\subsection{Inflation}

\textit{\emph{During a de Sitter epoch in which $H=const$, one has
$\mathcal{H}'\sim\mathcal{H}$ and from Eq.~(\ref{eq:rprime}): 
\begin{align}
r & \sim const\,\,.
\end{align}
The tensor equations (\ref{eq:gweq}) then reduce to: 
\begin{align}
h_{g}''+3h_{g}'+h_{g}(\frac{k^{2}}{\mathcal{H}^{2}}+\frac{a^{2}\beta_{1}r}{\mathcal{H}^{2}}) & =\frac{a^{2}\beta_{1}r}{\mathcal{H}^{2}}h_{f} \, \, , \\
h_{f}''+3h'_{f}+h_{f}(\frac{k^{2}}{\mathcal{H}^{2}}+\frac{a^{2}\beta_{1}}{r\mathcal{H}^{2}}) & =\frac{a^{2}\beta_{1}}{r\mathcal{H}^{2}}h_{g}
\end{align}
We can now assume $a^{2}\beta_{1}r\ll\mathcal{H}^{2}$ during inflation
(ie $\rho_{inf}\gg\rho_{mg}$) so $h_{g}$ behaves as in GR. The same
is true for $h_{f}$ since $a^{2}\beta_{1}/r\mathcal{H}^{2}\sim(\beta_{1}/\beta_{4})r^{-3}\ll1$
. Since the inflationary equations are the standard ones, we expect
the initial conditions to be unchanged and to apply equally well to
$h_{g}$ and $h_{f}$.}}

\subsection{Radiation and Matter Dominated Era}

\label{sec:tensors_radiation_matter}

In the early time, we can approximate the ratio of scale factors as
$r'=-\frac{3}{2}\left(1+w_{tot}\right)r$ which is solved by 
\begin{equation}
r=Aa^{-\frac{3}{2}\left(1+w_{tot}\right)},
\end{equation}
where $A$ is a suitable normalization constant of order unity. Furthermore
we approximate $\mathcal{H}^{2}\simeq\frac{1}{3}\beta_{4}a^{2}r^{2}$.
If the initial conditions for $h_{g}$ and $h_{f}$ are similar, then
the source terms (\ref{coeff4}) are negligible at early times, i.e.
small $a$, and the equations decouple. Furthermore, we find 
\begin{eqnarray}
\gamma_{g} & \simeq\frac{3}{2}\left(1-w_{tot}\right), & \qquad\gamma_{f}\simeq-\frac{3}{2}\left(3w_{tot}+1\right),\\
m_{g}^{2} & \simeq\frac{3\beta_{1}}{A\beta_{4}}a^{\frac{3}{2}(w_{tot}+1)}, & \qquad m_{f}^{2}\simeq-\frac{3\beta_{1}(3w_{tot}+1)}{2A^{3}\beta_{4}}a^{\frac{9}{2}(w_{tot}+1)},\\
c_{g}^{2} & \simeq1,\qquad\qquad\quad & \qquad c_{f}^{2}\simeq\frac{(3w_{tot}+1)^{2}}{4}.
\end{eqnarray}
Neglecting the mass term $m_{f}^{2}$ at early times, the tensor evolution
for $f_{\mu\text{\ensuremath{\nu}}}$ is described by \textit{\emph{ 
\begin{align}
h_{f}''-\frac{3}{2}(3w_{tot}+1)h_{f}'+\frac{h_{f}k^{2}(3w_{tot}+1)^{2}}{4\mathcal{H}^{2}} & =0.\label{eq:tensor_pert_approx_f}
\end{align}
At large scales the last term is negligible and one finds a growth
of $h_{f}$ as $a^{3\left(3w_{tot}+1\right)/2}$. Thus, when radiation
dominates, $h_{f}$ increases very fast as $a^{3}$.}} Clearly, if
one starts with $h_{f}'=0$ then this growing mode is initially absent
and it takes some time before it becomes visible. The evolution of
$h_{g}$ has instead a constant mode $h\sim const$ until $h_{f}$
is large enough to source the growth of $h_{g}$, cf Eq. (\ref{eq:coupling_early}).

The early time approximation that leads to Eq (\ref{eq:tensor_pert_approx_f})
turns out to be a very good approximation also in the matter domination.
In this regime $h_{f}$ increases as $a^{3/2}$ for super-horizon
modes. When the coupling term becomes important, $h_{g}$ is driven
by $h_{f}$ and acquires the same trend. Finally, when MDE ends and
the system approaches a de Sitter behavior, the perturbations begin
to decay.

For sub-horizon scales the behavior is influenced by the $h_{f}$
time-dependent sound speed. An asymptotic form for large $k$ can
however be found. In this regime we can neglect the mass and the coupling
terms, and the $h_{g},h_{f}$ equations during either RDE or MDE have
the general form 
\begin{equation}
h_{n}''+\gamma_{n}h_{n}'+\beta_{n}k^{2}a^{\eta}h_{n}=0 \, \, , 
\end{equation}
where the index $n$ stands for $g,f$ and $\eta=1+3w_{tot}$ and
$\beta_{n}$ is an irrelevant constant. The general solution can be
easily written in terms of the Bessel functions but here we need only
the asymptotic behavior for large $k$ or late times, which is 
\begin{equation}
h_{n}\sim a^{-(\frac{\gamma_{n}}{2}+\frac{\eta}{4})}
\end{equation}
times fast oscillations. We see then that for sub-horizon modes $h_{f}$
grows as $a^{1}$ in RDE, as $a^{1/2}$ in MDE and a final decay as
$a^{-1}$ when approaching the future deSitter phase, while $h_{g}$
decays as in the standard case as $a^{-1}$ in all eras (before being
driven to growth by the coupling to $h_{f}$). For very large wavenumbers
the coupling and the mass terms are ineffective at all times and the
$h_{g}$ equation reduces to the standard case. This implies that
there is no large effect to be expected for the directly detectable
range of gravitational waves, which is around 0.1Hz or $k\approx10^{14}$
Mpc$^{-1}$ (see e.g. \cite{Smith:2005mm}), although a precise calculation
is beyond the scope of this paper.

\begin{figure}
\begin{centering}
\includegraphics[width=0.5\columnwidth]{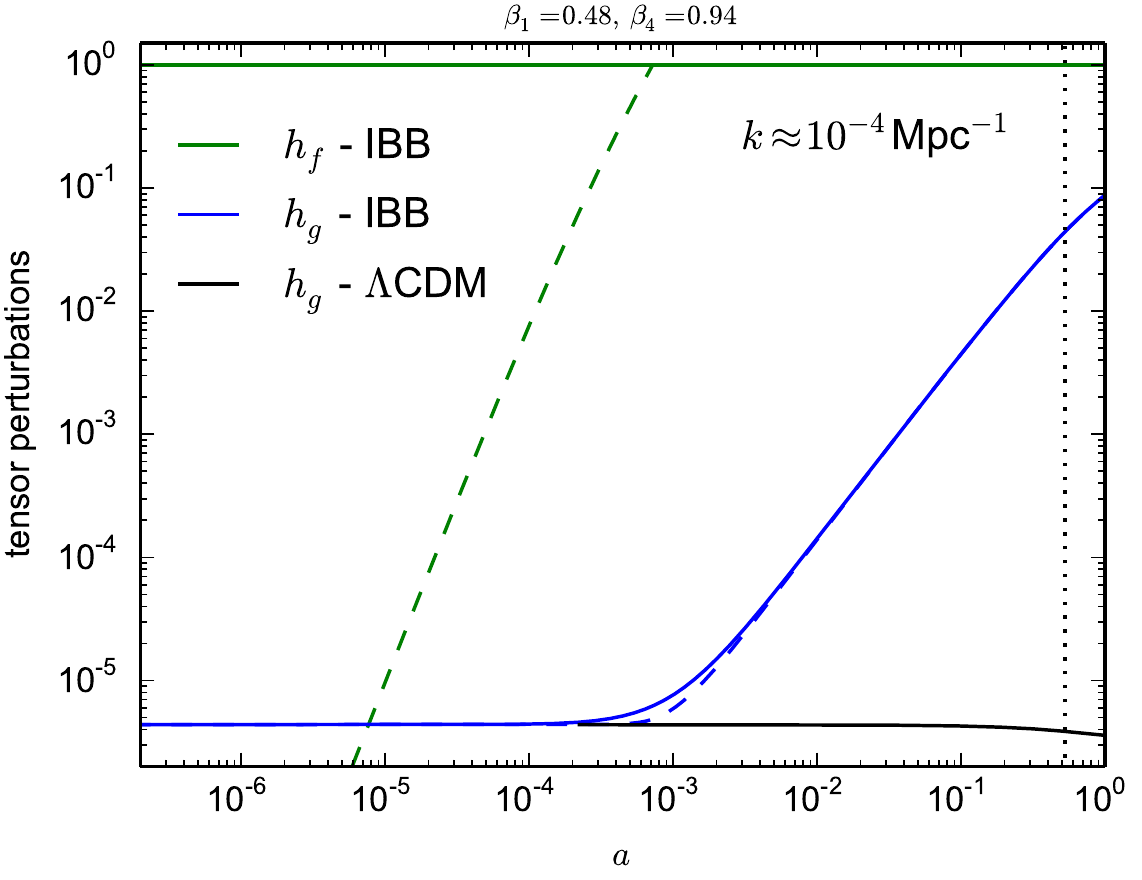}\includegraphics[width=0.5\columnwidth]{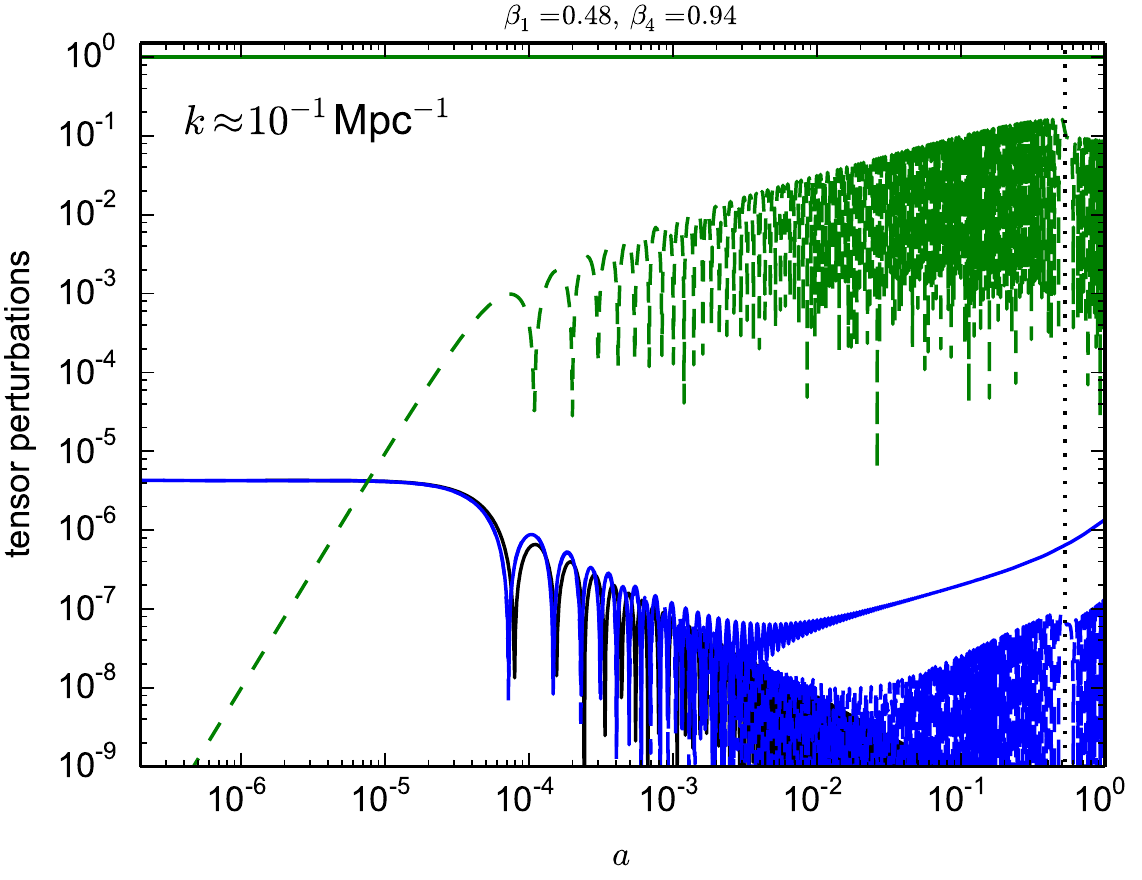}\\
 \includegraphics[width=0.5\columnwidth]{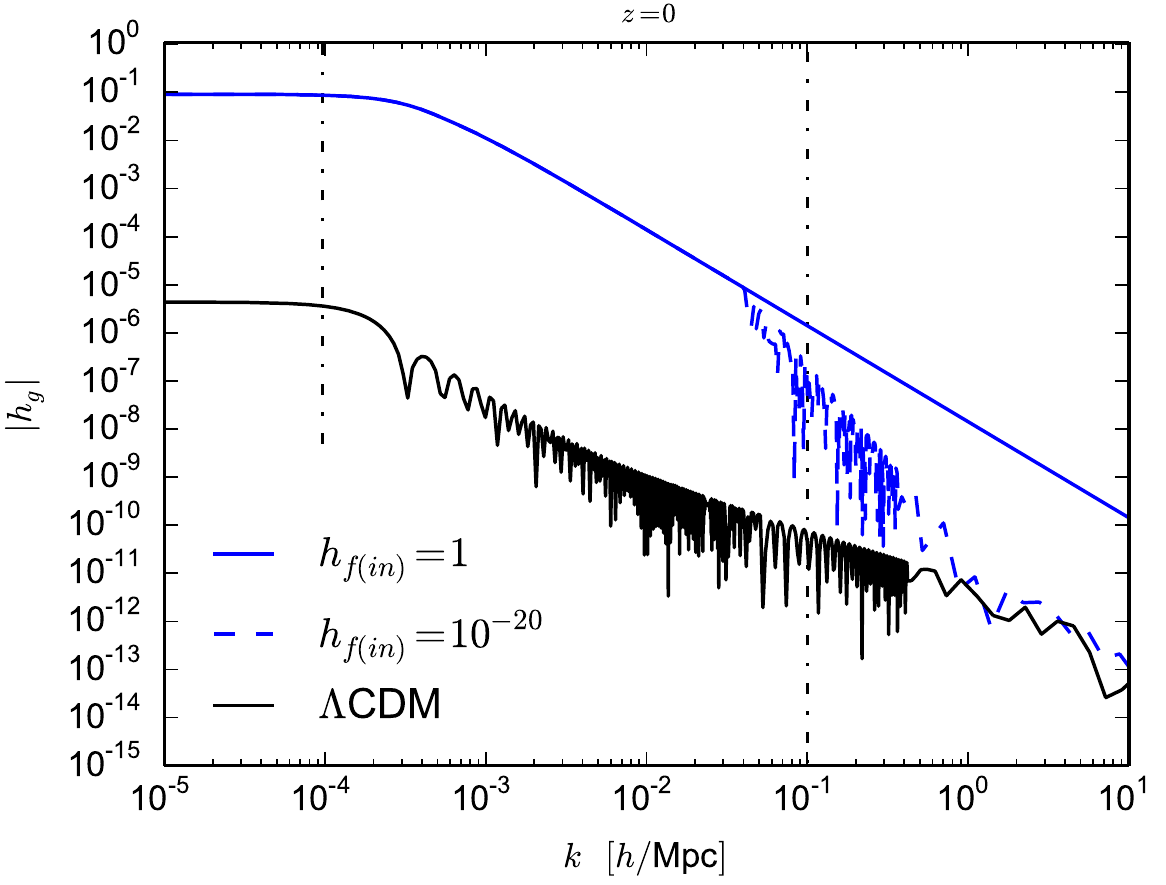}\includegraphics[width=0.5\columnwidth]{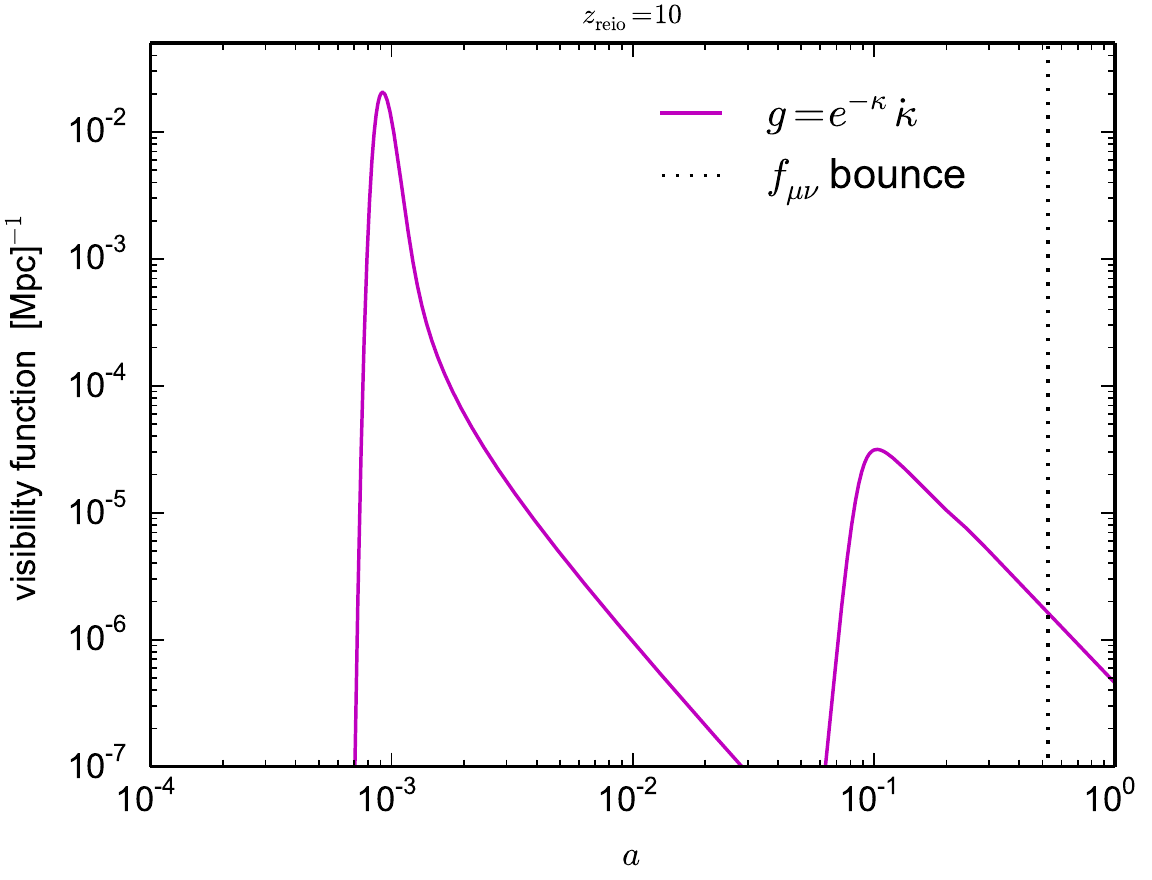} 
\par\end{centering}

\protect\protect\protect\protect\caption{ Evolution of tensor perturbations at different scales (top panels)
and scale dependence of tensor perturbations at $z=0$ (bottom left
panel). All scales include a cut-off when non-linearity is reached;
$h_{g}$ does not reach the cut-off for any scale and redshift. The
initial conditions have been chosen so that $h_{f}=1$ is initially
non-linear (solid lines) or starts at a small value (dashed lines)
and becomes non-linear at latter times. The latter choice corresponds
to $h_{f(in)}^{\prime}\approx h_{f(in)}=10^{-20}$ at $a=10^{-10}$
(see section \ref{sec:initial_conditions}). The two modes shown in
the top panel correspond to the vertical dash-dotted lines in the
bottom left panel. For reference, we recall the standard CMB photon
visibility function (bottom right panel), whose peaks correspond to
recombination and reionization epochs (see section \ref{sec:cmb}).
The bounce of the reference metric has been indicated with a dotted
vertical line. \label{fig:evolution}}
\end{figure}

\subsection{Inflationary initial conditions}

We can now use the $a^{3}$ growth mode during radiation to estimate
the order of magnitude effect of the tensor modes at recombination
(a more precise estimation will be obtained numerically in sections
\ref{sec:cmb}, \ref{sec:solutions}). Inflation ends at some energy
scale that can vary from $10^{15}$ GeV to few MeV depending on the
model. The upper limit comes from the bounds on the amplitude of tensor
perturbations, indicating that the energy scale of inflation is at
least that of Grand Unified Theories when observable modes are produced.
The lower bound is inferred from the need of a radiation dominated
universe in thermal equilibrium during big bang nucleosynthesis. These
values are reached when the scale factor was $a_{inf}\approx10^{-9}$
at the latest. Since super-horizon tensor modes grow as $a^{3}$ during
radiation domination, in the most favourable case of inflation ending
just before big bang nucleosynthesis one would obtain an enhancement
until recombination $a_{\mathrm{rec}}\approx10^{-3}$ of $h_{f(\mathrm{rec})}\sim10^{18}h_{f(e)}$
roughly, where the subscript $e$ denotes the end of inflation. If
$h_{f(e)}$ has the standard value approximately equal to $H_{e}/T_{P}\approx T_{e}^{2}/T_{P}^{2}$
during inflation, where $T_{P}\approx2.4\cdot10^{18}$ GeV is the
reduced Planck temperature/energy and $T_{e}$ is the inflationary
energy scale (here for simplicity assumed to be similar to the energy
at the end of inflation), then the value at recombination of $h_{f}$
for a wave that reenters horizon at recombination or larger is roughly
\begin{equation}
h_{f(\mathrm{rec})}\approx\left(\frac{a_{\mathrm{rec}}}{a_{e}}\right)^{3}h_{f(e)}=\left(\frac{T_{e}}{T_{\mathrm{rec}}}\right)^{3}\left(\frac{T_{e}}{T_{P}}\right)^{2}
\end{equation}
Shorter waves reenter before and therefore grow less. If this value
has to be compatible with the level of fluctuations in the CMB polarization
spectra, then it should be lower than about one tenth of the temperature
fluctuations; we take conservatively the level $10^{-7}$. The same
value should be taken for $h_{f}'$ since inflation excites both the
tensor mode and its momentum conjugate. However we do not detect directly
$h_{f}$ but rather the $g$-metric mode $h_{g}$ which is coupled
to matter, so as already noticed one can have a value of $h_{f}$
larger than $h_{g}$ by a factor of $q_{g}^{-1}\approx10^{4}$ at
recombination. Putting therefore $h_{f(\mathrm{rec})}\le10^{-3}$,
we obtain an upper limit to the temperature at the end of inflation
$T_{e}\approx10$ GeV. Since tensor modes impact CMB also at reionization,
when the coupling term $q_{g}$ is closer to unity, this limit should
be lowered to roughly 
\begin{equation}
T_{e}\approx1\,\,\mathrm{GeV}
\end{equation}
(a similar limit has been obtained also in \cite{Cusin:2014psa}).
It might be interesting to remark that the superhorizon
growing mode breaks the standard link between tensor modes and inflationary
scale due to the presence of the coupling: now in principle one can have observable tensor modes even
in low scale inflation. 

Any inflationary model with higher energy scale will generate excessive
power on the tensor modes unless the inflationary initial conditions
are suppressed with respect to the standard value or their growth
is reduced. Taken at face value, this shows that the $a^{3}$ growing
mode can be reconciled with observations only in the rather extreme
scenarios of very low-energy inflation, as e.g. in the models discussed
in Ref. (\cite{German:2001tz,Allahverdi:2006iq,Choudhury:2013jya}).
Fixing the initial conditions to a more conservative era for the end
of inflation, e.g. $T\sim10^{3}$ GeV, would produce the huge value
$h_{f(\mathrm{rec})}\sim10^{39}h_{f(e)}\sim10^{7}$.

Barring the case of very low-energy inflation, then, the IBB model
is at odds with CMB observations. In the rest of the paper we will
explore more or less contrived ways to overcome this difficulty.

\subsection{The non-linear cut-off}

\label{sec:cutoff}

The tensor perturbations in the reference metric grow so fast that
they will eventually become non-linear. At this point, the perturbative
treatment followed so far breaks down and one has to take into account
higher order corrections, or even the full equations of motion in
order to correctly reproduce the dynamics. A natural question is then
what happens to the two metrics after non-linearity is reached: the
evolution would then need to be calculated self-consistently in a
non-linear theory for bigravity, which is beyond the scope of this
paper. This problem is not new to Dark Energy models. There are cases
such as growing neutrino cosmologies \citep{Amendola:2007yx} in which
the effect of non-linearities becomes important and needs to be taken
into account also when dealing with the CMB predictions \citep{Pettorino:2010bv}.
In that scenario, the fast growth actually leads to stable non-linear
structures (which are a way to test the model rather than an argument
to exclude it based on linear theory).

Following the idea of Ref. \cite{Pettorino:2010bv}, we then stop
the evolution of perturbations at some cut off amplitude value $h^{cut}\approx1$,
when non-linearity is approximately reached. This prescription is
applied to both $h_{g}$ and $h_{f}$ and has the effect to partially
stop the `dragging' of the second metric $h_{f}$ over the standard
one $h_{g}$. Such assumption is adopted here for simplicity, as for
now our interest is to give a consistent estimate of how big is the
impact of the growing mode on CMB spectra and tensor perturbations
when non-linearity is reached.

In practice, since the growth of $h_{f}$ is very rapid in the early
Universe, this is equivalent to fixing $h_{f}=h^{cut}=1$ from the
very beginning, with the consequence that the perturbations of the
reference metric are not dynamical anymore. Nevertheless they still
affect the tensor modes of the physical metric due to the coupling
(\ref{eq:coupling_early}). The overall evolution as a function of
the scale factor for two values of $k$ and two different choices
of the initial conditions is plotted in Fig.{\ref{fig:evolution}}
for both $h_{g}$ and $h_{f}$, once the bound has been applied. Figure
\ref{fig:evolution} also shows a similar behavior between a model
in which $h_{f}$ starts saturated and one in which the cutoff value
is reached during the evolution. It is also shown how, even though
the cutoff is applied to both metrics, $h_{g}$ never reaches it
during its evolution.
Note that the full non-linear dynamics might produce other effects.
For example, in the limit of scales smaller than the horizon, one
finds the oscillating behavior in e-folding time $h_{g},h_{f}\sim e^{imN}$
with eigenfrequencies: 
\begin{equation}
m=\pm\frac{r+r'}{\mathcal{H}r}.\label{eq:oscillatory_behavior}
\end{equation}
This oscillating behavior is then always present in the sub-horizon
solutions, overimposed to an amplitude modulation, as shown for the
smallest scale in Fig. \ref{fig:evolution} (top right panel). In
this case, setting $h_{f}$ to the non-linear cut-off value leads
to a growing behavior plus a damping of the initial oscillations.
On the other hand, the model with fine tuned initial conditions displays
the oscillatory behavior expected from linear theory (\ref{eq:oscillatory_behavior}),
and in this case the non-linear value $h_{f}\sim1$ is not reached,
at least for Fourier modes corresponding to small scales. In this
case the negative friction of the reference metric gets compensated
by the positive friction from the physical metric. Nonetheless we
expect our method to give a correct quantitative estimate of the observable
effects of the growing mode on the CMB.

\section{Cosmic Microwave Background anisotropies in bigravity}

\label{sec:cmb}

The Cosmic Microwave Background (CMB) was shown to be a powerful probe
to test not only early time cosmology but also Dark Energy and Modified
Gravity models \cite{Ade:2015rim}. In particular, in this paper we
are interested in the effect that tensor perturbations in bigravity
have on the CMB power spectra. At recombination, when photons are
not anymore tightly coupled to baryons but decoupling has not occurred
yet, electrons can be scattered simultaneously by photons coming from
cold and hot spots. In presence of a quadrupole temperature anisotropy,
the scattered photons will be linearly polarized and the CMB radiation
will be characterized not only by its intensity, but also by its polarization.
CMB polarization can be expressed in a tensor normal basis in Fourier
space, in terms of E and B modes. While scalar perturbations can only
produce an E mode (primordial) polarization pattern, tensor perturbations
can feed both E and B primordial modes. Therefore any change in the
evolution of tensor perturbations predicted in bigravity will affect
the polarization spectra.

At later times, polarization and temperature anisotropies are further
modified during reionization. Reionization occurs at a much lower
redshift, when the universe becomes partially ionized due to the formation
of the first stars, allowing CMB photons to partially rescatter. The
recombination and reionization eras correspond to peaks in the visibility
function shown in figure \ref{fig:evolution}. The visibility function
$g(t)=\exp(-\kappa)\dot{\kappa}$ (where $\kappa$ is the optical
depth and $\dot{\kappa}$ is its derivative with respect to conformal
time $\tau$) gives the probability that a photon last scattered in
the conformal time interval $[\tau,\tau+d\tau]$. Due to the importance
of the coupling at relatively low redshift, the most important effects
of tensor modes on the CMB are imprinted during the reionization epoch.

In the following, we have only modified tensor perturbations, assuming
that the contribution of scalar perturbation is small enough to be
neglected, as scalar modes affect B mode polarization only indirectly,
via lensing of E modes, at scales $\ell\gtrsim150$. Of course, if
polarization is large enough, it might also feed back the scalar spectra.
However, this seems a good enough first approximation to test the
specific effect of the growing mode on the BB spectra. We implemented
the tensor evolution equations in two publicly available Boltzmann
codes, CAMB \cite{Lewis:2002ah} and CLASS \cite{Blas:2011rf}, and
compared the results obtained in the various cases to verify their
mutual consistency. 

\begin{figure}
\begin{centering}
\includegraphics[width=0.5\columnwidth]{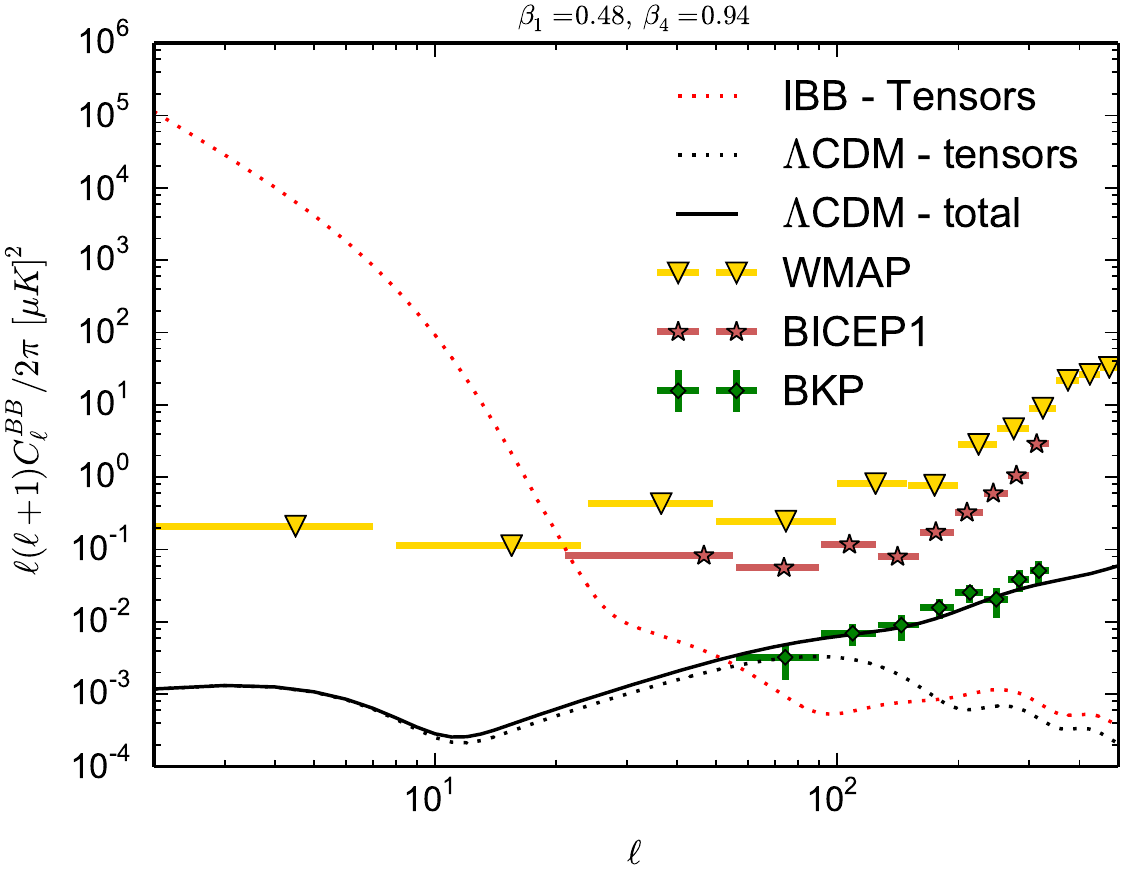}\includegraphics[width=0.5\columnwidth]{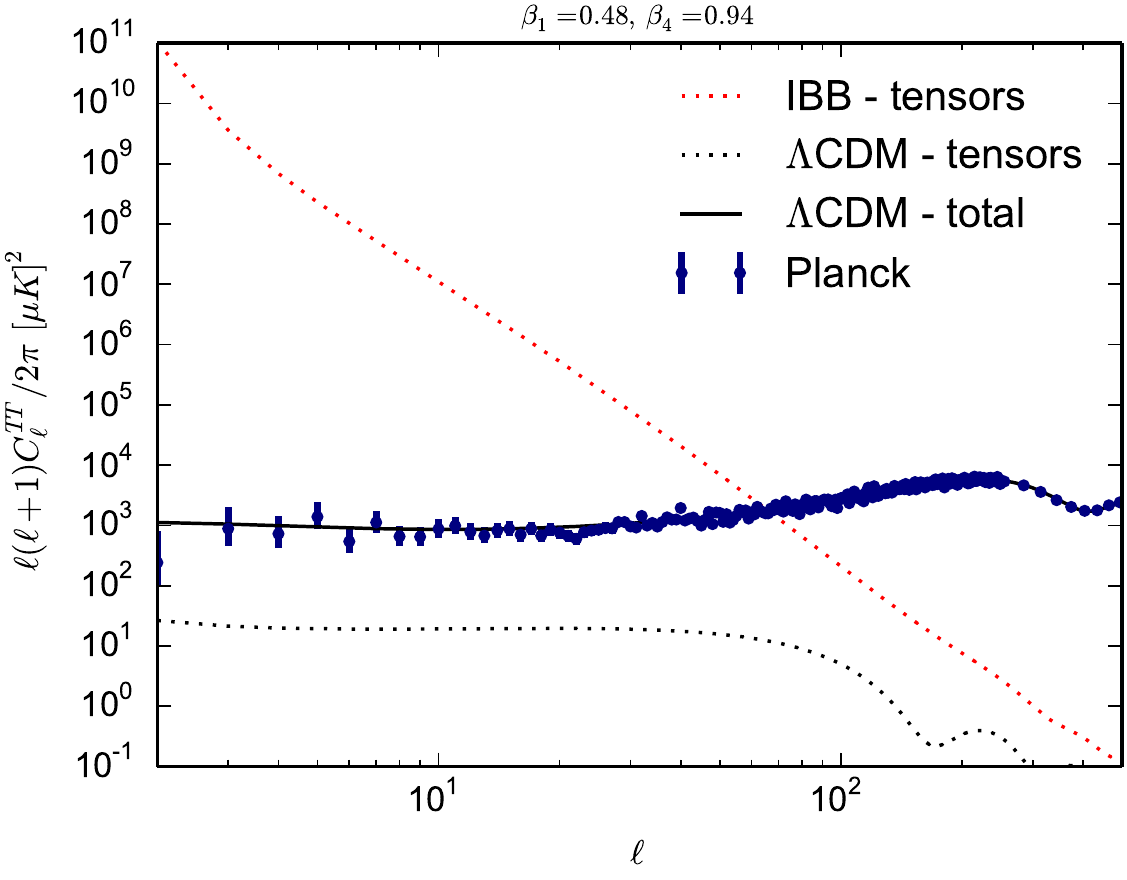}
\includegraphics[width=0.5\columnwidth]{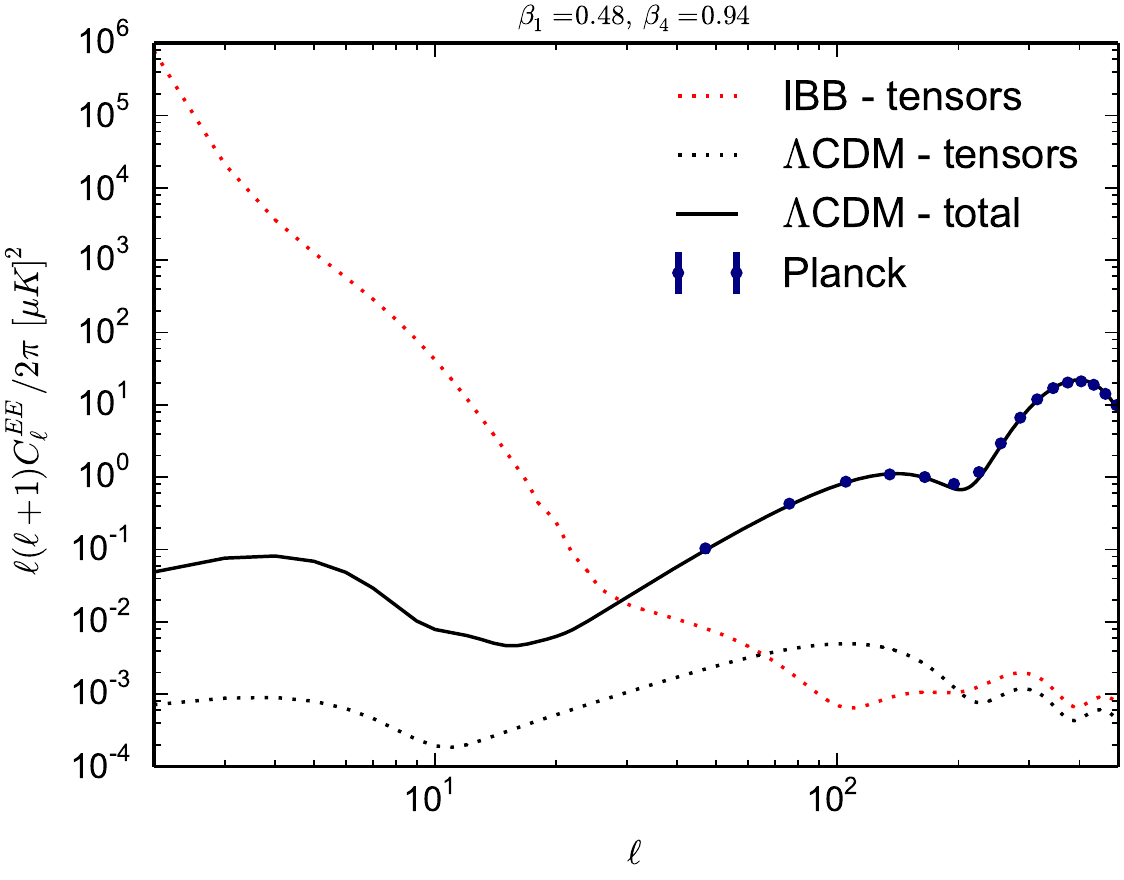}\includegraphics[width=0.5\columnwidth]{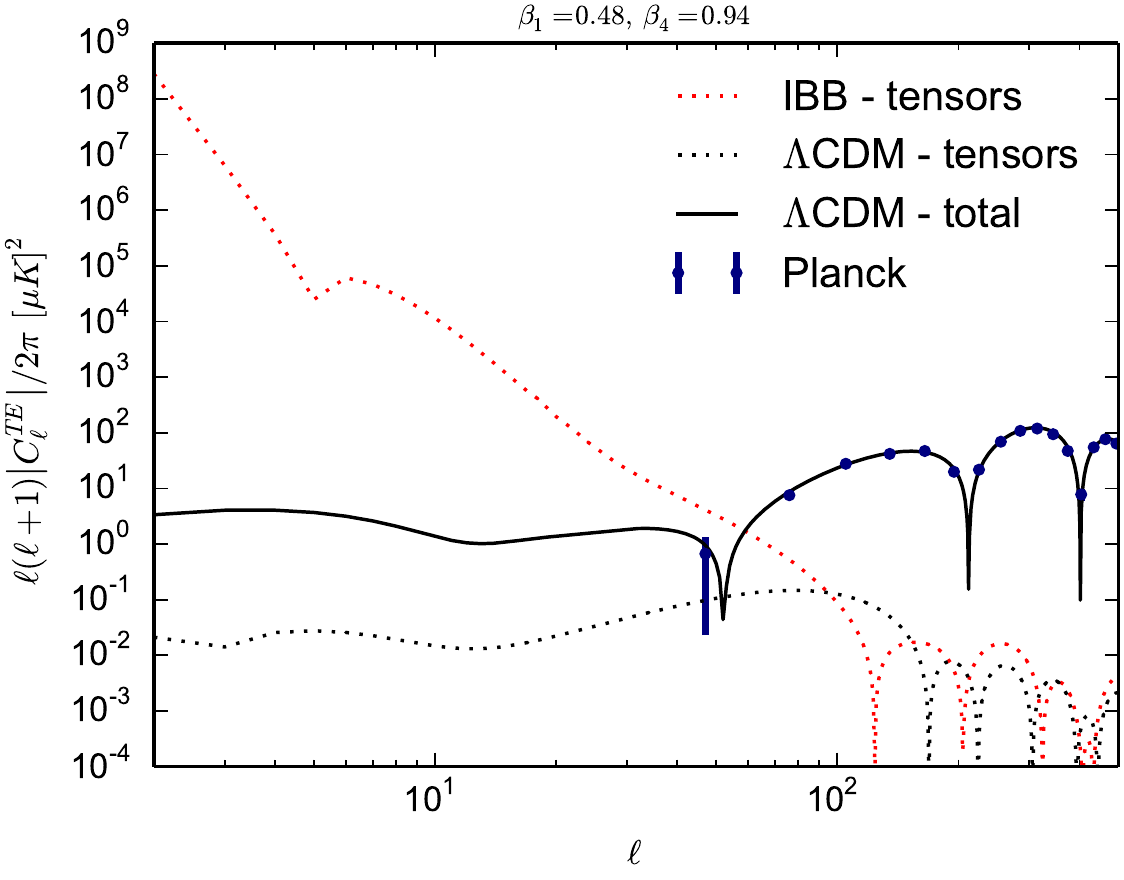} 
\par\end{centering}

\protect\protect\protect\protect\caption{Effects of the growing mode in the $f$-tensors on the CMB. All the
plots assume IBB with $\beta_{1}=0.48$, $\beta_{4}=0.94$, $\Omega_{m}=0.18$
with $\Omega_{b}=0.05$, $h=0.67$, $r_{T/S}=0.05$ and the spectral
index determined by inflationary self-consistency conditions. The
tensor perturbations have been assumed to start at non-linear cutoff
value $h_{f}^{{\rm }}=1$. \label{fig:spectra_fiducial} }
\end{figure}

As discussed in the previous section, our aim is to check the effect
on the CMB spectra consistently, i.e. taking into account that, by
definition, we cannot trust any result derived assuming linear perturbation
theory when perturbations become non-linear. We then fix $h_{f}=h^{cut}=1$
from the beginning and evolve only the $g$-metric tensor $h_{g}$,
for which we will assume standard initial conditions: 
\begin{equation}
A_{t}=r_{T/S}A_{s}\left(\frac{k}{k_{p}}\right)^{n_{T}}\,,
\end{equation}
with a fiducial tensor-to-scalar ratio $r_{T/S}=0.05$, a scalar amplitude
$A_{s}=2.21\times10^{-9}$ and a tensor spectral index given by the
self-consistency condition of single field slow-roll inflation $n_{T}=-(2-r_{T/S}/8-n_{s})r_{T/S}/8$,
where the fiducial scalar spectral index is $n_{s}=0.9645$ (in section
\ref{sec:initial_conditions} we explore the effects of changing the
IC on the tensor modes.). For bigravity we choose the best fit model
$\beta_{1}=0.48$, $\beta_{4}=0.94$ with $\Omega_{cdm}=0.13$, $\Omega_{b}=0.05$,
while for the reference $\Lambda$CDM model we choose $Planck$ 2015
TT, EE, TE+lowP marginalized values \citep{Planck:2015xua}, i.e.
$\Omega_{cdm}=0.26$, $\Omega_{b}=0.05$. In both bigravity and $\Lambda$CDM
cases the fiducial optical depth is $\tau=0.079$, corresponding to
$z_{reio}=10$. In order to test the IBB model, we compare the achieved
spectra with up to date CMB observations, using Planck 2015 data \citep{Adam:2015rua}
for TT, TE and EE spectra, while for the BB spectrum we rely on WMAP
\citep{2012arXiv1212.5225B} and BICEP1 \citep{Barkats:2013jfa} together
with the joint BICEP2, Keck, Planck analysis \citep{Ade:2015tva}.
Figure \ref{fig:spectra_fiducial} shows the tensor contribution to
the CMB temperature, polarization and cross spectra for the fiducial
bigravity model described above. Large angular scales are the most
sensitive to the growing modes in the $f$-metric, because\textcolor{blue}{{}
}\textcolor{black}{the coupling to the physical metric is only important
at low redshifts, after recombination. On these scales the tensor
perturbations give contributions to the power spectra that are too
large to be compatible with CMB data.} Even for the T and E polarization
spectra, the tensor contribution in IBB bigravity overshoots the \textcolor{black}{observed
values by several orders of magnitude }for $\ell\lesssim60$. Since
both the scalar and tensor contributions to TT and EE spectra are
positive definite, it is impossible that a reduced scalar contribution
compensates the (large) tensor part in order to fit the data. The
conclusion is that the cut-off on the growing mode is not enough to
save the model.

\section{Possible solutions to the growing mode problem}

\label{sec:solutions}

\begin{figure}
\centering{}\includegraphics[width=0.5\columnwidth]{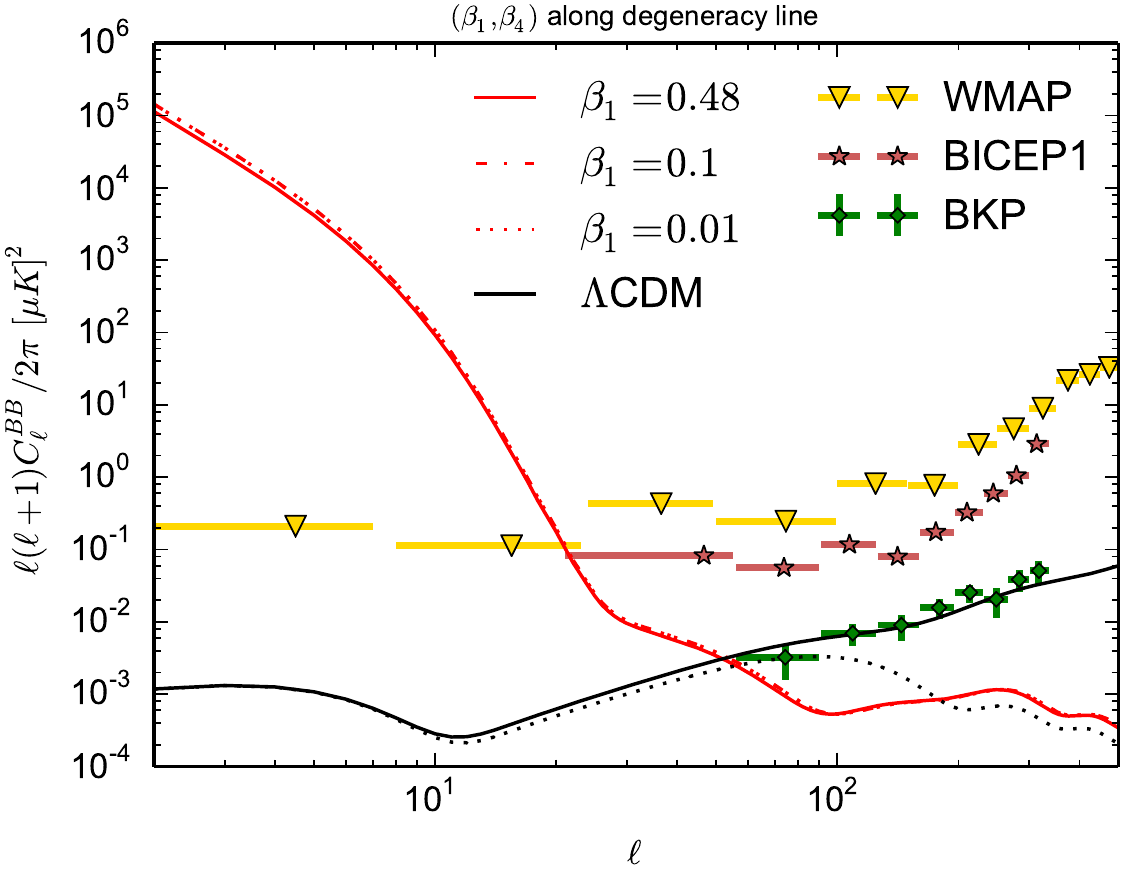}\includegraphics[width=0.5\columnwidth]{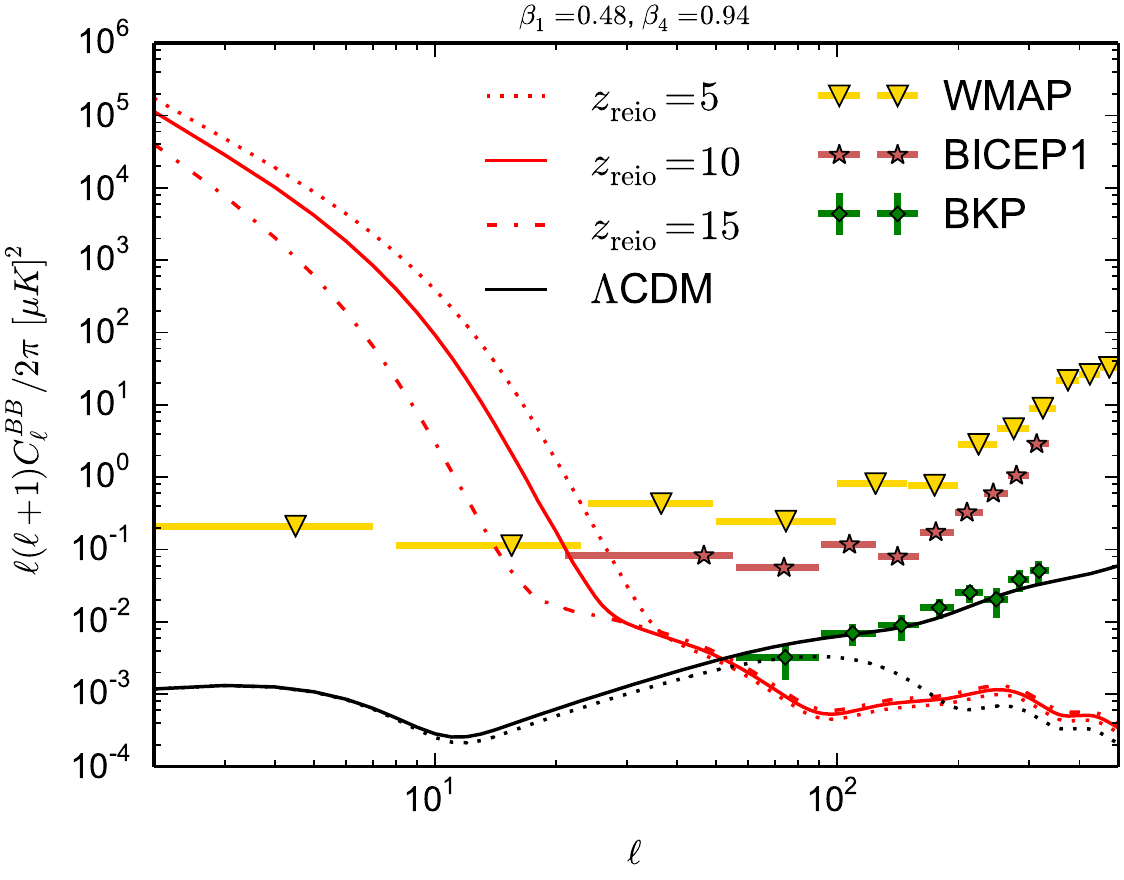}
\includegraphics[width=0.5\columnwidth]{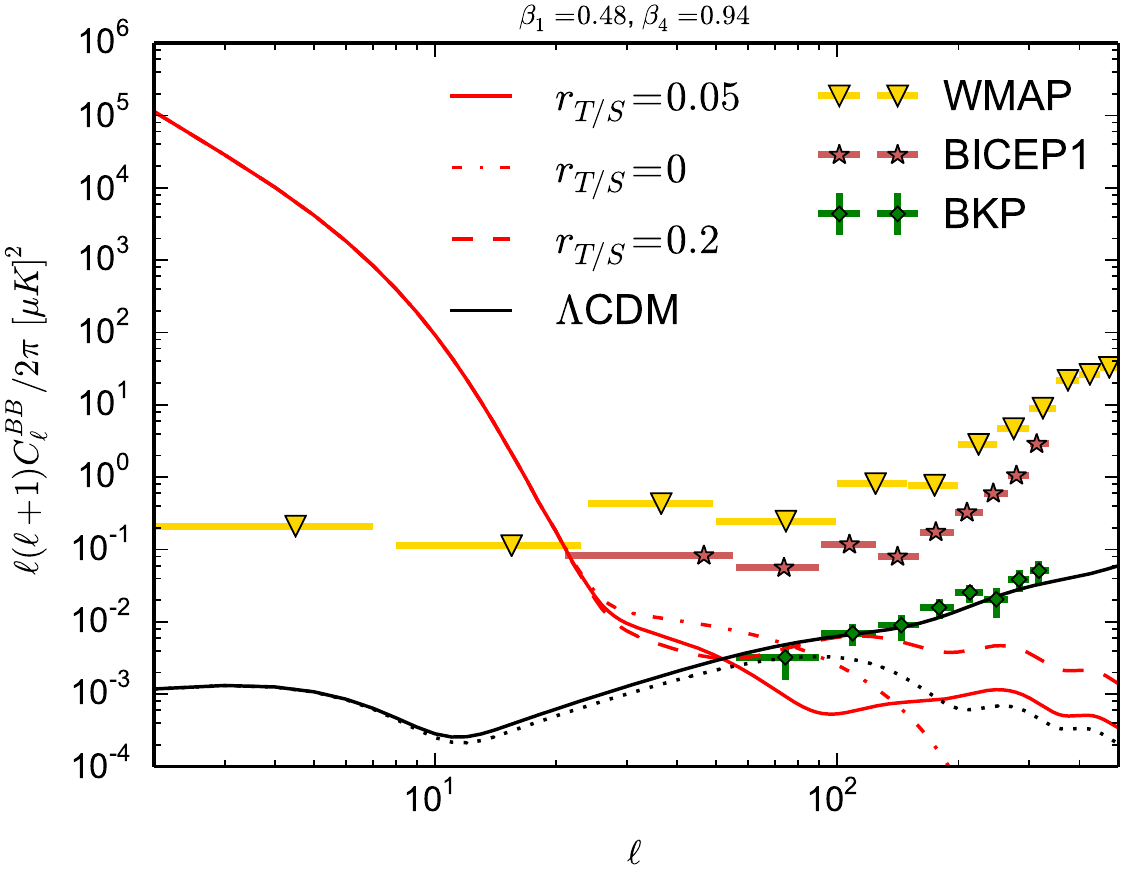}\includegraphics[width=0.5\columnwidth]{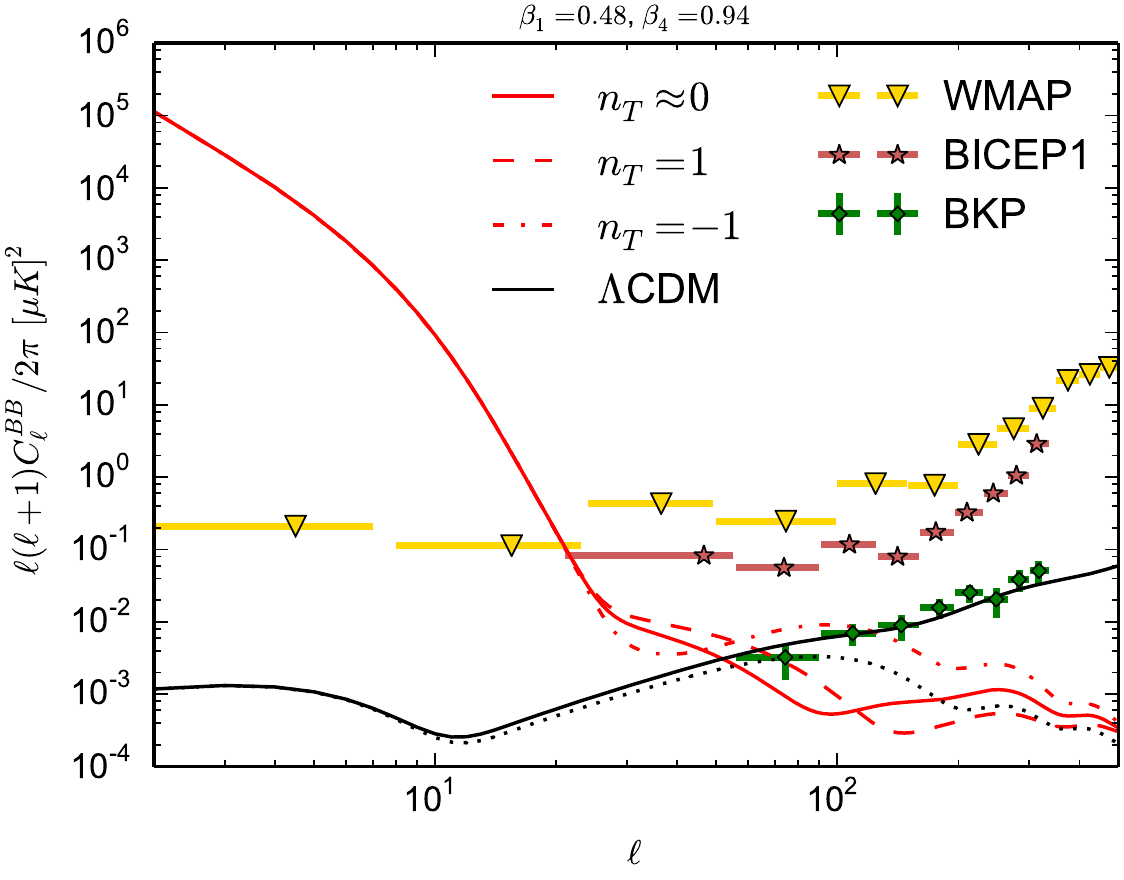}
\protect\protect\protect\protect\caption{Non-solutions to the problem of growing modes in the reference metric:
varying the IBB parameters across the degeneracy line (top left panel),
the reionization redshift (top right) and the initial conditions for
tensor perturbations in the physical metric (bottom panels). The remaining
model details are the same as in Fig. \ref{fig:spectra_fiducial}.
In particular, the reference metric perturbations are set initially
to the non-linear cutoff value. All solid lines correspond to the
standard values described in section \ref{sec:cmb}. \label{fig:non-solutions}}
\end{figure}

A first attempt to overcome the growing mode problem is to verify
how much the effect on the spectra depends on the choice of the fiducial
model. We investigated the effect of changing the bigravity parameters
$\beta_{1}$ and $\beta_{4}$, choosing them close to the degeneracy
curve (\ref{eq:deg_curve}), the redshift of reionization $z_{reio}$,
the tensor to scalar ratio $r_{T/S}$ and the tensor spectral index
$n_{T}$ characterizing the shape of the primordial power spectrum
for tensor perturbations. These effects are shown in Fig. \ref{fig:non-solutions},
showing that simple variations of these cosmological parameters do
not offer a sufficient improvement.

The observable impact on the CMB is produced in the reionization era
because the coupling between the physical and reference metric's tensor
perturbations is only relevant at low redshift. Shifting $z_{{\rm reio}}$
in the range $(5,15)$ has only a small effect on the BB spectrum.
Further changes would spoil the predictions for EE and TE spectra
and enter in contradiction with the Gunn-Peterson limit \cite{Becker:2001ee}.
Higher reionization redshifts reduce the tension slightly, mainly
because the perturbations in $h_{g}$ are smaller at earlier times
and on smaller scales. Another attempt that does not work is to modify
the initial conditions for perturbations of the physical metric. Varying
the spectral index and the tensor to scalar ratio only has an impact
on relatively high multipoles $\ell>30$, on which the tensor perturbations
are predominantly imprinted during recombination. For lower multipoles
the evolution of $h_{g}$ is dominated by the coupling to $h_{f}$,
which overshoots the initial conditions on large scales due to its
large value.

Concordance with observations can neither be achieved by varying the
IBB parameters $\beta_{1},\beta_{4}$. The fast growth of the $h_{f}$
tensor perturbations leads to growing $h_{g}$ tensors due to the
coupling in (\ref{eq:gweq}) which is proportional to $\beta_{1}r/\mathcal{H}^{2}$.
One might expect that a change in the betas could lead to higher values
of $r$ and, thus, a suppression of the coupling at early times (note
that $\mathcal{H}^{2}\propto r^{2}$ in RDE). Higher values of $r$
are possible when lowering $\beta_{1}$ since $r_{0}=\mbox{\ensuremath{\left(1-\Omega_{tot0}\right)}/}\beta_{1}$,
where $r_{0}$ denotes the present value of $r$ and one has $r>r_{0}$
during the entire evolution. However, in order to fit observations
we have to choose parameters being close to the degeneracy curve (\ref{eq:deg_curve}),
i.e. $\beta_{4}/\beta_{1}^{2}\simeq\text{const}$. As already mentioned,
at early times the coupling term is then proportional to $q_{g}\approx1/\beta_{1}r$
(see Eq. \ref{eq:couplqg}). A smaller coupling would of course help
in delaying or reducing the effect of the $h_{f}$ growing mode on
$h_{g}$. Since the evolution for large $r$ is nearly independent
of the coupling parameters (see Eq. \ref{eq:early}), and since $\beta_{1}\apprle0.5$,
the ratio $1/(\beta_{1}r)$ cannot decrease much below the reference
case. Figure \ref{fig:non-solutions} shows clearly that the effect
of varying $\beta$'s is negligible, at least when the coefficients
remain along the degeneracy line.

One loophole in the line of argument above is to
leave the observational degeneracy curve (\ref{eq:deg_curve}) and
add an explicit cosmological constant $\beta_{0}$. This transforms
the IBB model into a form of $\Lambda$CDM plus a small admixture
of massive bigravity. The coupling coefficient $q_{g}$ does not explicitly
depend on $\beta_{0}$ and reduces to: 
\begin{equation}
q_{g}=\frac{3\beta_{1}r^{2}}{\beta_{1}+\beta_{4}r^{3}}.
\end{equation}
From the $f_{\mu\nu}$-Friedmann eq. (\ref{eq:hubble}) we find that
$r$ scales like $\beta_{4}^{-1/2}$ for small values of $\beta_{1}$
which leads to: 
\begin{equation}
q_{g}\simeq\frac{9\beta_{1}}{\beta_{1}\beta_{4}+3\sqrt{3\beta_{4}}}.
\end{equation}
Thus, one is able to get an arbitrarily small coupling when choosing
values of $\beta_{1}$ that are sufficiently smaller than $\beta_{4}$.
In this regime, the massless graviton dominates over the massive one
and the cosmological evolution tends to that of $\Lambda$CDM. It
is clear that such a model is not particularly interesting from a
cosmological point of view because hardly distinguishable from the
standard model and therefore here we will not investigate it further.

In order to render the model viable without adding a cosmological
constant, it is necessary to adopt a more radical solution. In the
following subsections we explore how IBB can be reconciled with observations
by fine-tuning the initial conditions in the reference metric perturbations,
lowering the non-linear cut-off or modifying the theory.

\subsection{Changing the initial conditions for $h_{f}$}

\label{sec:initial_conditions}

\begin{figure}
\centering{}\includegraphics[width=0.5\columnwidth]{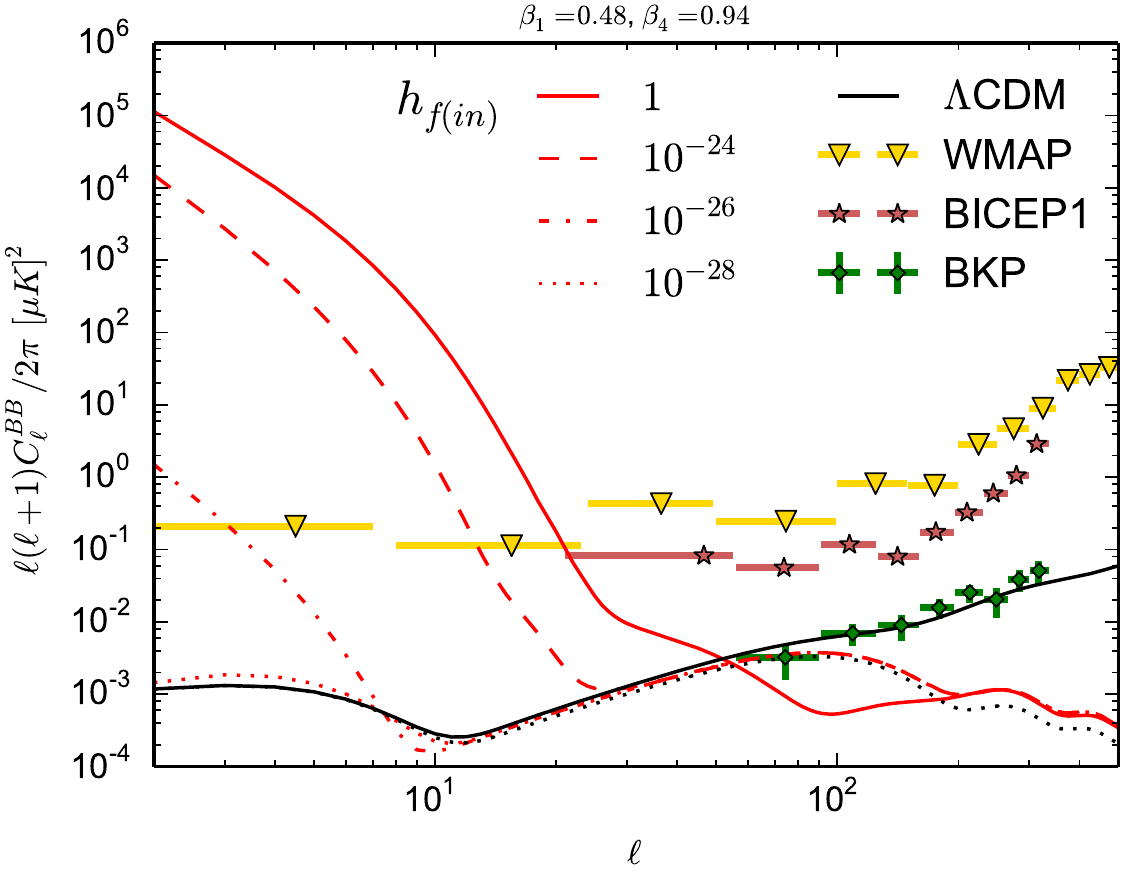}\includegraphics[width=0.5\columnwidth]{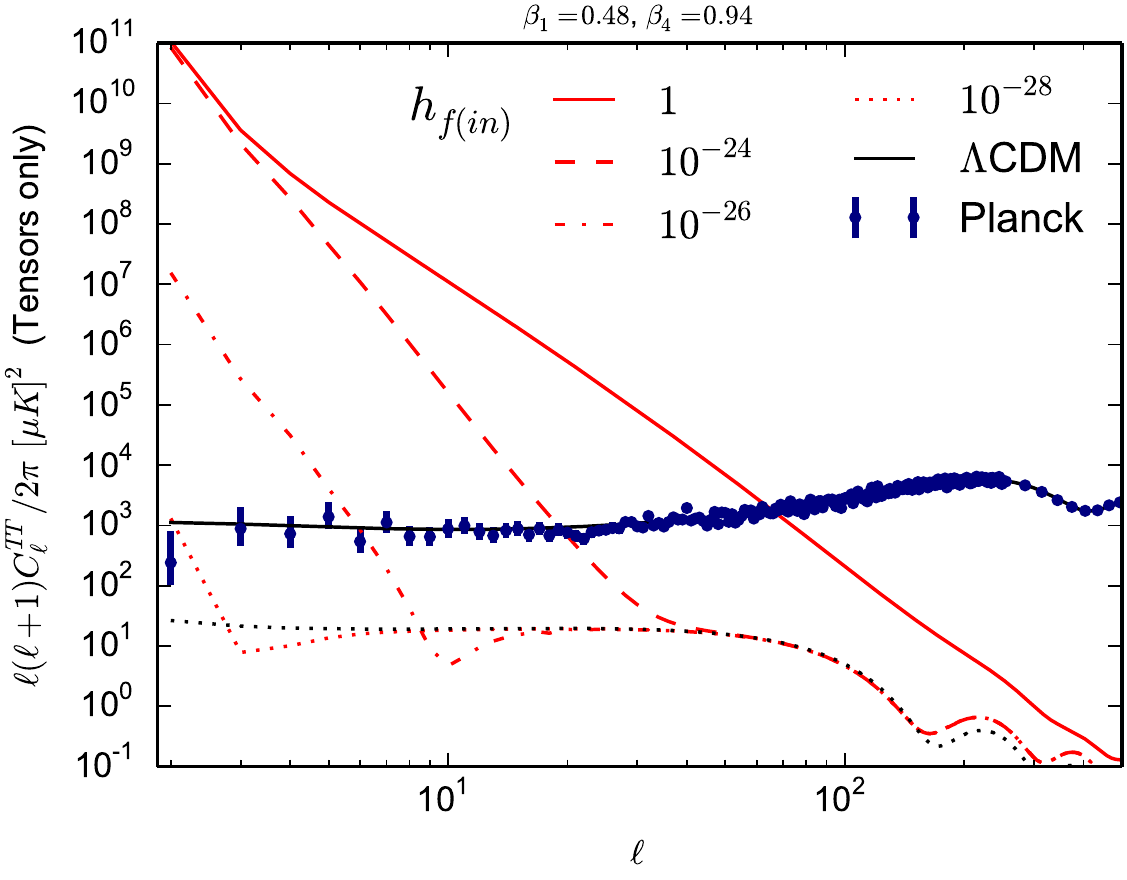}
\protect\protect\protect\protect\protect\protect\protect\protect\protect\caption{BB (left panel) and TT (right panel) spectra using fine-tuned initial
conditions as described in section \ref{sec:initial_conditions}.
The initial amplitude and time derivative have been specified according
to the growing solution (\ref{eq:growing_ic}) at a fiducial scale
factor $a_{{\rm in}}=10^{-10}$. Bigravity BB spectra (red lines)
only contain the primordial tensor contribution, while $\Lambda$CDM
spectra (black solid lines) includes the contribution of both scalar
and tensor perturbations. The evolution of tensor perturbations has
been stopped whenever $h_{g,f}=1$ is reached (cf. section \ref{sec:cutoff}).
\label{fig:fine_tuned_IC}}
\end{figure}

Our next attempt consists in checking whether fine tuning the initial
conditions can compensate the effect of the growing mode on the CMB
spectra, as illustrated in Fig.(\ref{fig:fine_tuned_IC}). We specify
the initial conditions in terms of the growing solution in the radiation
era 
\begin{equation}
h_{f}(a)=h_{f(\mathrm{in})}\left(a/a_{{\rm in}}\right)^{3}\,,\quad h_{f}^{\prime}=3h_{f}(a)\,,\label{eq:growing_ic}
\end{equation}
found in section \ref{sec:tensors_radiation_matter}.%
\footnote{One can in general fix $h_{f}$ and $h_{f}^{\prime}$ independently
for each wavenumber, but we restrict to the simpler choice (\ref{eq:growing_ic})
here. If more general IC are considered, a necessary condition for
the growing mode to be sufficiently suppressed is that the time derivative
is small. This condition is sufficient as long as $h_{f({\rm in})}$
is well below the cutoff value (see next subsection).%
} We find that the initial conditions have to be fine tuned to zero
to at least the level of one part in $10^{26}$ at $z_{\mathrm{in}}=10^{10}$
in order to fit current limits on the BB spectrum. This choice of
the initial epoch corresponds to an era before Big Bang Nucleosynthesis,
which as already mentioned is a hard lower bound for the end of inflation.
One can easily relate to earlier times in order to specify the IC
at the reheating epoch, when inflation ends and tensor perturbations
start growing. Table \ref{tab:bounds_tuned_IC} extrapolates the result
to the range of energies in which inflation might have ended.

\begin{table}[h!]
\begin{tabular}{l@{\qquad}c@{\qquad}c@{\qquad}c@{\qquad}c}
 & BBN  & fiducial  & 1 GeV  & GUT \tabularnewline
\hline 
Bounds on $h_{f({\rm in})}$  & $\lesssim10^{-19}$  & $\lesssim10^{-25}$  & $\lesssim10^{-31}$  & $\lesssim10^{-82}$ \tabularnewline
Scale factor $a$  & $\sim10^{-8}$  & $\sim10^{-10}$  & $\sim10^{-12}$  & $\sim10^{-29}$ \tabularnewline
Temperature $T$  & $\sim0.1$MeV  & $\sim10$MeV  & $1$GeV  & $\sim10^{16}$Gev \tabularnewline
\hline 
\end{tabular}\protect\protect\protect\protect\protect\protect\protect\caption{Upper bounds on $h_{f}$ extrapolated to different epochs using the
growing mode, eq. (\ref{eq:growing_ic}). The limits shown are based
on the results for BB spectra, which only depend on the tensor sector
at low multipoles where the theory enters in tension with the data.
Considering TT spectra would tighten these bounds, as can be inferred
from the left panel of figure \ref{fig:fine_tuned_IC}. \
\label{tab:bounds_tuned_IC}}
\end{table}

As already noticed, the only way to generate naturally a very low
level of tensor modes compatible with CMB without any cut-off is to
assume that inflation ends at an energy scale not larger than 1 GeV.
If however one assumes the non-linear cut-off, then the inflation
energy scale bound can be relaxed. From Table \ref{tab:bounds_tuned_IC}
we see empirically that the initial condition for $h_{f}$ is related
to the end of inflation energy scale $T_{e}$ (expressed in GeV) as
$h_{f({\rm in})}\approx10^{-31}T_{e}^{-3}$. Then one has 
\begin{equation}
h_{f({\rm in})}\approx\left(\frac{T_{e}}{T_{P}}\right)^{2}\approx10^{-31}\left(\frac{1GeV}{T_{e}}\right)^{3}\,,
\end{equation}
or $T_{e}\approx25$ GeV, a more realistic scale range for low-energy
inflation.

A sufficiently small value of $h_{f}^{\prime}$ might also be provided
by a more exotic inflationary mechanism. During inflation no growing
modes occur on the perturbations of the $f$ metric. Some solutions,
such as increasing the mass of the graviton at very early times, might
naturally generate the low values needed to reconcile the model with
observations (the problem of growing classical perturbations is common
to ekpyrotic scenarios alternative to inflation \cite{Kallosh:2001ai}).
Outside of these rather unconventional, although not impossible, cases,
the conclusion we draw is that only very fine tuned initial conditions
allow to reconcile bigravity with CMB observations.

Even if inflation ends at a sufficiently low scale or a mechanism
to suppress $h'_{f({\rm in})}$ exists, it has been argued by Cusin
et al. \cite{Cusin:2014psa} that non-linear corrections would spoil
the small value of $h_{f}^{\prime}$. Although we will not investigate
this issue further here, we note that the nature of the theory might
protect the tensor modes against such terms. This is precisely what
happens in the linear equation (\ref{eq:gweq}), in which the source
term is highly suppressed. If a similar suppression occurs also on
the non-linear source terms, the fine tuned initial conditions can
render the model compatible with CMB observations (assuming that there
are no additional complications in the scalar sector).

\subsection{Lowering the cut-off}

\begin{figure}[t]
\centering{}\includegraphics[width=0.5\textwidth]{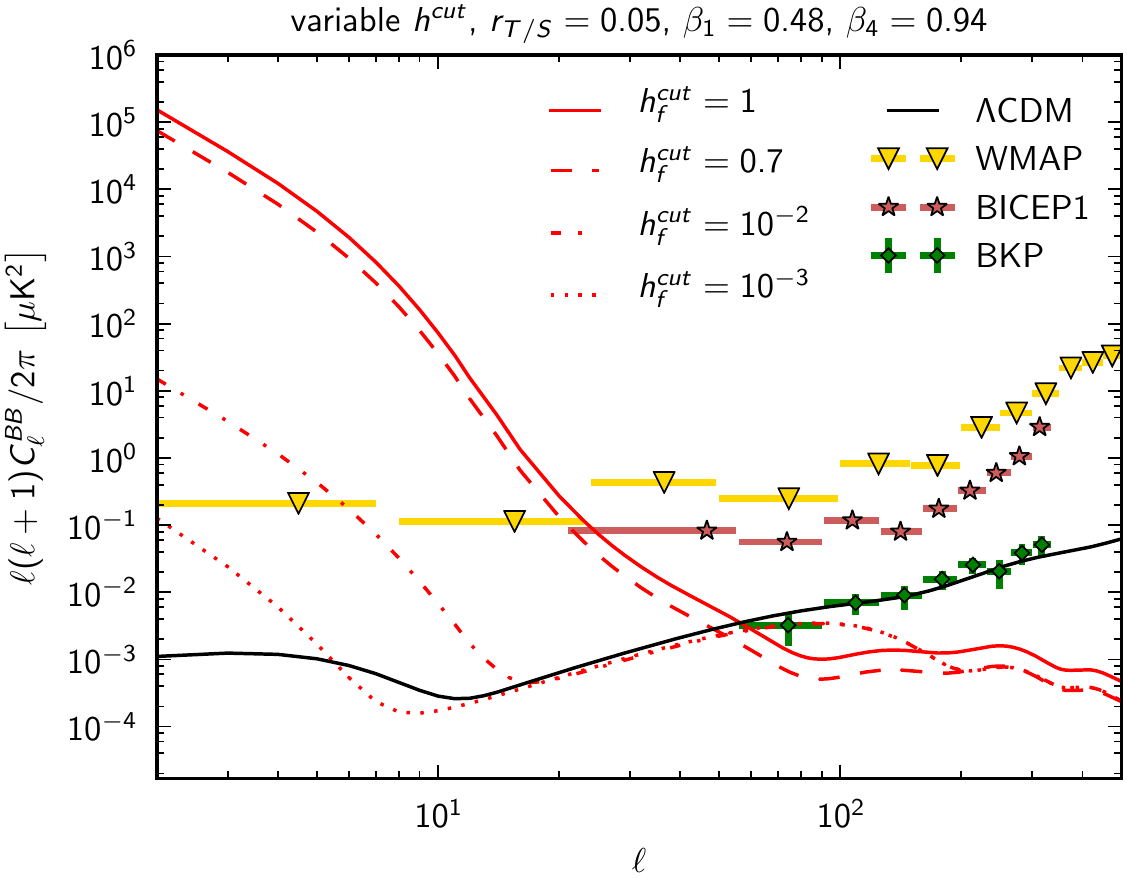}\includegraphics[width=0.5\textwidth]{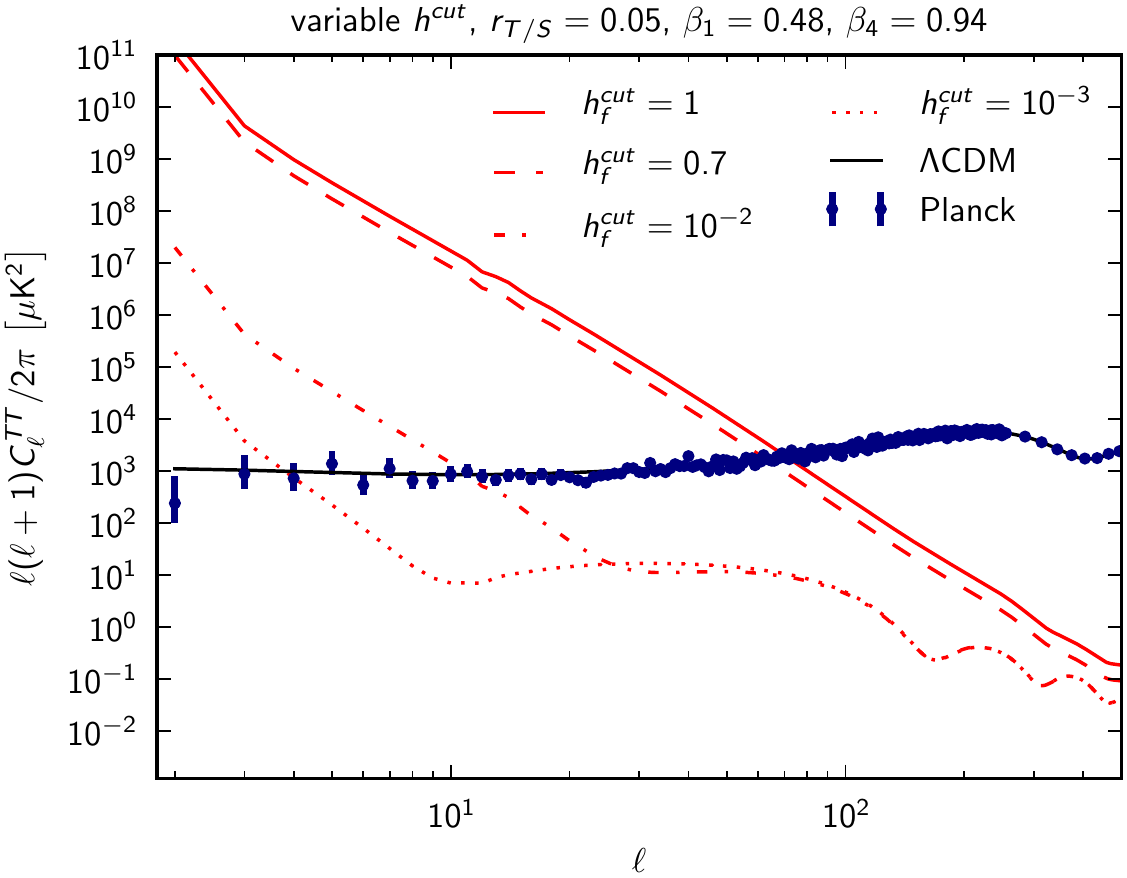}
\protect\protect\protect\protect\caption{BB (left panel) and TT (right panel) using a varying cutoff $h^{cut}$.
The evolution of $h_{f}$ and $h_{g}$ is frozen when those reach,
respectively, the cutoff value $h^{cut}$. Bigravity BB spectra (red
lines) only contain the primordial tensor contribution, while $\Lambda$CDM
spectra (black solid lines) and observational data points contains
the contribution of both scalar and tensor perturbation. \label{fig:lowcut} }
\end{figure}

Another possible solution to mitigate the effect of IBB on CMB spectra
is to assume that non linear effects begin to be not negligible before
$h_{f}$ reaches unity; this can be thought as an effective way to
treat the impact of non linear effects which, even if dominant when
$h_{f}$ is above unity, can start to affect the evolution of the
perturbations even for lower values. Therefore, in Figure \ref{fig:lowcut}
we show the behavior of TT and BB spectra for different values of
$h^{cut}$ at which we freeze the evolution of the metric perturbations.
One would need to suppress the cutoff scale by at least three orders
of magnitude to reconcile theoretical BB spectra with currently available
limits and possibly an even lower value to make the TT spectrum acceptable
for current data, once the scalar contribution is taken into account.

It seems contrived that non-linear effects might play a role at such
small values of the cutoff. However, theories of massive gravity are
known for having strong non-linear effects in certain limits, such
as the Vainshtein mechanism \cite{Vainshtein:1972sx,Babichev:2013usa}.
As long as numerical results for the non-linear evolution of tensor
perturbations are not available, we must contemplate the possibility
that the cut-off could be lower than unity and even significantly
lower.

\subsection{Modifying the theory}

The growing modes in the reference metric could possibly be reconciled
with CMB data by a suitable modification of the theory. Here we explore
a phenomenological modification in which a redshift dependence of
the $\beta_{i}$ parameters is assumed; this kind of behavior might
be achieved in generalized massive gravity models \citep{deRham:2014gla},
where a time dependence of the mass parameters is introduced without
the addition of any new dynamical degree of freedom. Our modification
consists on setting $\beta_{1}=\beta_{4}=0$ until a certain switch
redshift $z_{s}$ is reached. Then bigravity becomes active and the
evolution described in Section \ref{sec:tensor_perturb} is switched
on. The considered background evolution instead is the one produced
by bygravity at all redshift, as IBB well approximates the standard
background, which should take place at $z>z_{s}$, at early times.

\begin{figure}[t]
\centering{}\includegraphics[width=0.5\textwidth]{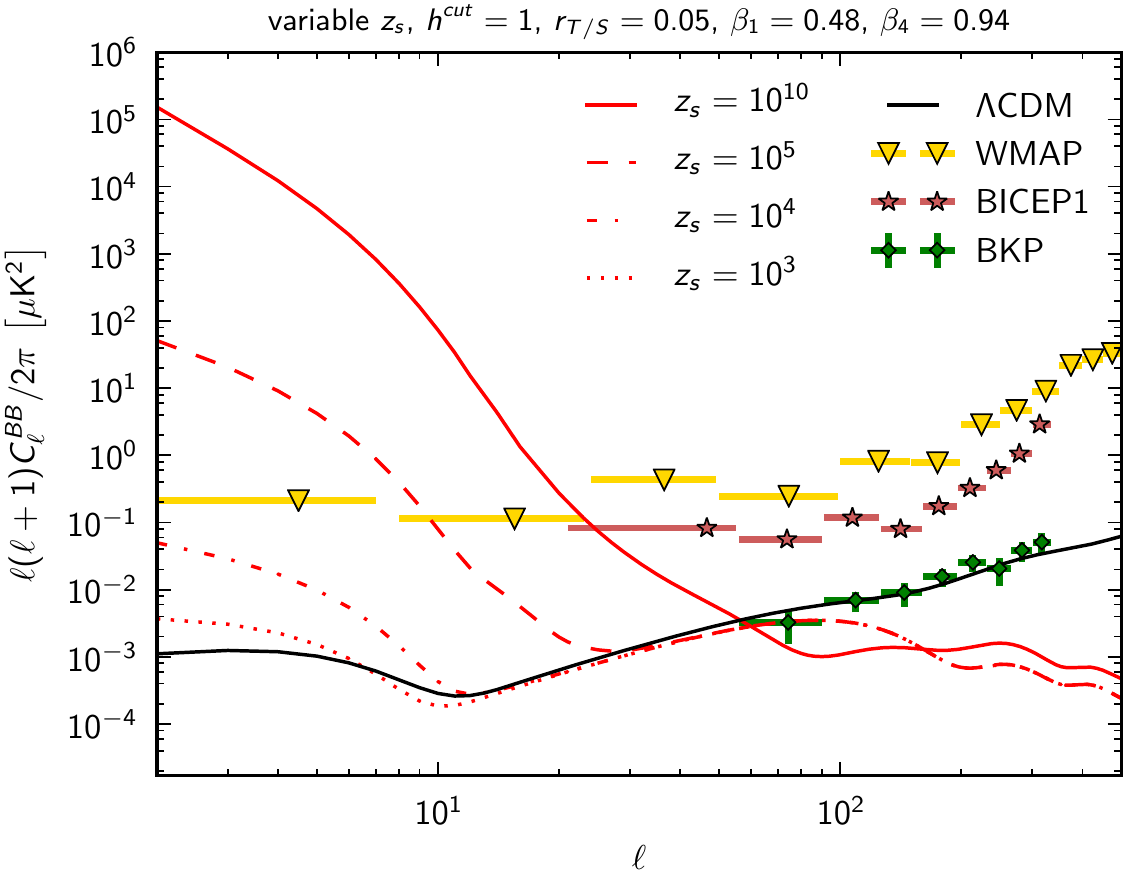}\includegraphics[width=0.5\textwidth]{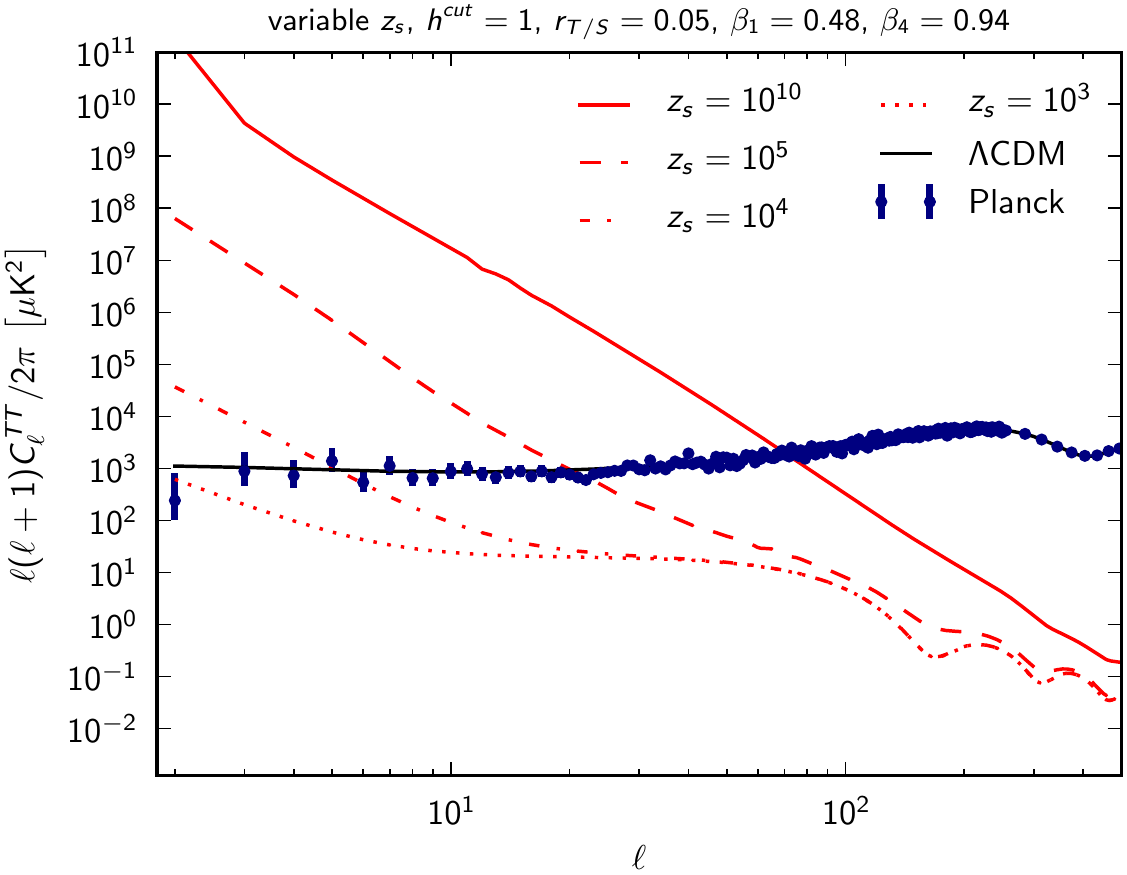}
\protect\protect\protect\protect\protect\caption{BB (left panel) and TT (right panel) using a varying initial redshift
$z_{s}$. If $z>z_{s}$ $h_{f}$ follows the same standard equation
as $h_{g}$, while for $z\leq z_{s}$ the evolution described in Section
\ref{sec:tensor_perturb} is switched on. Bigravity BB spectra (red
lines) only contain the primordial tensor contribution, while $\Lambda$CDM
spectra (black solid lines) and observational data points contains
the contribution of both scalar and tensor perturbation. \label{fig:redvar} }
\end{figure}

The results in Figure \ref{fig:redvar} show how switching on bigravity
at approximately the redshift of matter-radiation equality ($z_{s}\approx10^{3}$)
can produce an acceptable BB spectra when comparing with current data.
In the TT case instead, the contribution of scalar modes to the spectrum
can possibly lead to the necessity of an even lower value of $z_{s}$.

There are additional ways in which the theory might be modified while
retaining the original field content of bigravity. In the following
we will describe these possibilities, although addressing them in
detail will be left for future work. A possible modification is to
allow for branches different than IBB. So far, only branches in which
$r$ evolves from $r=0$ or $r\rightarrow\infty$ were considered.
In \cite{1475-7516-2014-03-029} all remaining cases due to a non-viable
behavior were excluded. However, some of those conditions based on
expectations of a standard cosmological evolution, like an expansion
at all times and the existence of a matter/radiation dominated era
in the asymptotic past. It might be interesting to study these disregarded
branches. Additionally, both metrics are usually assumed to be FLRW
at background level, even though this is assumed only for simplicity.
In Ref. \cite{Nersisyan:2015oha} the authors considered more general
types of metrics which might lead to interesting evolutions at background
and linear level and could also have an impact on the tensor evolution.

An additional possibility is to modify the coupling to matter. Even
though both metrics are a priori equally footed, we let only one of
them couple to matter while the remaining metric stays unobservable.
An additional coupling of the reference metric to matter would influence
the tensor perturbation and might be able to tame the fast growth
(see e.g. Refs. \cite{2013JCAP...10..046A,Akrami:2014lja} for further
discussion on bi-metric couplings) . An additional coupling of the
same matter Lagrangian to $f_{\mu\nu}$ is not possible as it will
generally reintroduce the BD ghost \cite{deRham:2014naa,2014arXiv1408.0487Y,2014arXiv1408.5131N}.
Even though this will not happen if a different matter Lagrangian
(an unobservable dark sector) is coupled to $f_{\mu\nu}$, we will
usually meet a new fine tuning \cite{Aoki:2014cla} that would make
the theory less appealing. One way out would be a coupling through
a new composite metric that is constructed such that it avoids the
BD ghost \cite{deRham:2014naa}. This choice would lead to viable,
self-accelerating backgrounds \cite{Enander:2014xga} but still does
not yield a realistic cosmological evolution at the linear level \cite{Gumrukcuoglu:2015nua,Comelli:2015pua}.

\section{Conclusions}

We have analyzed the behavior of tensor perturbations in the infinite-branch
bigravity (IBB) model and the signatures they produce on the CMB.
In this model the reference metric contracts at early times, causing
tensor perturbations in this metric to grow rapidly. These modes have
ample time to grow since the end of inflation. However, a consistent
analysis of linear perturbations and CMB spectra cannot include the
regime in which perturbations become non-linear. We then assume that
this growth stops when perturbations become non-linear, with an amplitude
saturated at a value of order unity. The coupling between the two
metrics produces in turn a growth of the tensor perturbations in $g$.
If the coupling is weak enough at early times, the growing mode will
in principle propagate to the physical metric only after some time.
Our first objective has been then to check whether this effect is
late enough to keep the spectra of tensor perturbations compatible
with present CMB data. Our conclusion is that, even when perturbations
remain below order unity, they are still large enough to have a twofold
impact on CMB spectra: first, the tensor modes provide a large contribution
to the TT, EE and TE spectra which is orders of magnitude larger than
the scalar contribution; second, the tensor modes induce a strong
B-mode polarization on the CMB. Both effects dominate on the largest
angular scales and are incompatible with observations from CMB experiments.

Varying the IBB parameter ($\beta_{1}$, with $\beta_{4}$ being derived
from it via Eq.(\ref{eq:deg_curve})) or other cosmological parameters
offers little help in reconciling IBB bigravity theory with observations.
We further explore five scenarios in which the theory might be rendered
viable: 
\begin{enumerate}
\item Lowering the energy scale of inflation. If the energy scale of inflation
is very small, around 1 GeV, the tensor modes are naturally suppressed
\emph{and} the growing mode has less time to grow until recombination
or reionization: the combined effect makes the IBB model acceptable
without any change. If one invokes the freezing of the growing mode
when it reaches non-linearity, then the inflationary energy scale
can be increased up to 25 GeV roughly. Such a scale is much lower
than the one predicted in simple slow roll inflation but could in
principle be achieved in alternative scenarios \cite{Kallosh:2001ai}. 
\item Fine-tuning the initial conditions. If $h_{f}^{\prime}$ is very small
at early times, the perturbations will not have reached the non-linear
value at late times, when the coupling to $g$ becomes important.
This requires a fabulous degree of fine-tuning, of one part in $10^{26}$
at $z=10^{10}$ ($10^{28}$ when TT modes are considered). For this
solution to work beyond the linear approximation, it is necessary
that non-linear sources in the equation for $h_{f}$ are suppressed
at early times. Even if stable agains non-linear corrections, fine
tuning the initial conditions for $h_{f}$ seems a highly ad-hoc requirement
for the theory in the absence of a mechanism, perhaps generalizing
inflation, able to naturally produce such small initial conditions.
As mentioned in the previous point, this fine-tuning occurs naturally
only in the case of very low-energy inflation. 
\item Lowering the non-linear cutoff. IBB becomes safe if $h_{f}$ stabilizes
at a value smaller than unity due to non-linear effects. This requires
the non-linear effects to act at most when $h_{f}\lesssim10^{-3}$
(and $\lesssim10^{-4}$ when TT modes are considered). 
\item Adding a cosmological constant. In that case the
theory still describes a massive graviton, although it would not be
responsible for the acceleration of the universe. From a purely cosmological
point of view, the model will be very similar to $\Lambda$CDM. 
\item Modifying the theory. One possibility is to allow for a time dependence
to the theory parameters $\beta_{1},\,\beta_{4}$ in the tensor perturbation
equations. This phenomenological parametrization of modified IBB allows
to satisfy CMB data if the parameters become non-zero only after $z=1000$. 
\end{enumerate}
While our assumption stops the growth of perturbations when they become
non-linear, a full analysis would require an understanding of the
actual non-linear behaviour of perturbations, which could be used
to exclude or validate such scenarios in a fully consistent way.

Finding a modification of bigravity that overcomes the difficulties
of the growing modes could be possible within the framework of generalized
massive gravity \cite{deRham:2014gla}. In this class of theories,
the interaction terms are given an additional dependence in the Stückelberg
fields, which allows the couplings to vary over time without introducing
additional degrees of freedom. Another possibility in this direction
is allowing a composite coupling that involves both tensor, possibly
on an equal footing. Theories with additional degrees of freedom,
such as scalar-bitensor or multigravity, might as well prove useful
to solve the problem of growing modes. More exotic modifications of
the theory remain to be explored.

Constructing a viable theory of massive gravity has proven to be a
challenge. Only after eight decades could the linear Fierz-Pauli theory
be generalized to a ghost-free non-linear completion, albeit one that
forbids any interesting cosmological solution. This difficulty could
be overcome by giving a kinetic term to the reference metric, allowing
the existence of accelerating cosmologies at the price of two additional
degrees of freedom, corresponding to a massless tensor. Of all the
five-parameter set of bigravity theories, only the two-parameter IBB
family is able to accelerate the universe with neither a cosmological
constant nor scalar instabilities. Yet, such a theory is affected
by growing modes that generically spoil the predictions of the cosmic
microwave background. The results presented here represent a setback
for the simple and appealing self-accelerating bigravity paradigm, a
paradigm that, unless saved by non-linear effects or a tiny amplitude
of the initial conditions, will have to be abandoned.
\begin{acknowledgments}
We thank Yashar Akrami, Ruth Durrer, Pedro Ferreira, Daniel G. Figueroa,
Fawad Hassan, Macarena Lagos, Gerasimos Rigopoulos, Angnis Schmidt-May,
Adam Solomon and Christof Wetterich for useful discussions. This research
has been supported by DFG through the grant TRR33 ``The Dark Universe''.
F.K. acknowledges support from the Graduate College ``Astrophysics
of fundamental probes of gravity''. 
\end{acknowledgments}
\bibliography{massive-gravity,all}

\end{document}